# Increasing robustness of pairwise methods for effective connectivity in Magnetic Resonance Imaging by using fractional moment series of BOLD signal distributions

Natalia Z. Bielczyk[a,b,*] , Alberto Llera[a,c], Jan K. Buitelaar[a,b],

Jeffrey C. Glennon[a,b], Christian F. Beckmann[a,b,c,d]

a) Donders Institute for Brain, Cognition and Behavior, Kapittelweg 29, 6525 EN Nijmegen, the Netherlands

b) Department of Cognitive Neuroscience, Radboud University Nijmegen Medical Centre, Geert Grooteplein Zuid 10, 6525 GA Nijmegen, the Netherlands

c) Radboud University Nijmegen, Comeniuslaan 4, 6525 HP Nijmegen, the Netherlands

d) Oxford Centre for Functional MRI of the Brain, University of Oxford, Oxford OX3 9DU, United Kingdom

---

**Abstract**

Estimating causal interactions in the brain from functional magnetic resonance imaging (fMRI) data remains a challenging task. Multiple studies have demonstrated that all current approaches to determine direction of connectivity perform poorly even when applied to synthetic fMRI datasets. Recent advances in this field include methods for pairwise inference, which involve creating a sparse connectome in the first step, and then using a classifier in order to determine the directionality of connection between of every pair of nodes in the second step. In this work, we introduce an advance to the second step of this procedure, by building a classifier based on fractional moments of the BOLD distribution combined into cumulants. The classifier is trained on datasets generated under the Dynamic Causal Modeling (DCM) generative model. The directionality is inferred based upon statistical dependencies between the two node time series, e.g. assigning a causal link from time series of low variance to time series of high variance. Our approach outperforms or performs as well as other methods for effective connectivity when applied to the benchmark datasets. Crucially, it is also more resilient to confounding effects such as differential noise level across different areas of the connectome.

---

## 1 Introduction

In the context of functional Magnetic Resonance research, effective connectivity refers to the process of estimating causal interactions between distinct regions within the brain. Several characteristics of fMRI data impose severe restrictions on the possibility of estimating such effective connectivity (Valdes-Sosa, Roebroeck, Daunizeau, & Friston, 2011, Friston, 2011, Bielczyk et al., 2018). First, the temporal resolution of the image acquisition is low (sampling rate typically < 1[Hz]). Furthermore, blood oxygen-level dependent (BOLD) activity is delayed with respect to neuronal firing, with a delay of 3-6[s] in the adult human brain (Arichi et al., 2012). The delayed hemodynamic response can also induce spurious cross-correlations between two BOLD time series (Ramsey et al., 2010, Bielczyk, Llera, Buitelaar, Glennon, & Beckmann, 2017). Both subject-to-subject and region-to-region variability in the shape of hemodynamic response (Devonshire et al., 2012) provide a general limitations to the methods for effective connectivity research in fMRI: when the hemodynamic response in one region is faster than in another, the temporal precedence of the peak of the hemodynamic response can easily be mistaken for causation. Secondly, fMRI data is characterized by a relatively low signal-to-noise ratio. Within grey matter and at field strengths of 3 [T], the task-induced signal changes range within 2-3% of the mean signal depending on the task (Kruger, 2018). Furthermore, the stochastic noise in the brain has been shown to have a scale-free spectral characteristic (He, 2014, Bédard, Kröger, & Destexhe, 2006, Dehghani, Bédard, Cash, Halgren, & Destexhe, 2010) which additionally hinders identifiability of causal structures derived from fMRI data (Bielczyk, Llera, Buitelaar, Glennon, & Beckmann, 2017). Moreover, typical fMRI protocols involve a relatively short time series (a few hundred samples), in which estimation of conditional probabilities between variables, and higher order statistic in the time series becomes difficult. Multiple methods were proposed to estimate effective connectivities from fMRI data (Friston, 2011). In the computational study by Smith et al. (Smith et al., 2011), a range of methods for effective connectivity were tested on synthetic datasets derived from a Dynamic Causal Modeling forward model (DCM, Friston, Harrison, & Penny, 2003). In this study, most methods for estimating causal interactions remained at chance level. One method highlighted as relatively successful at identifying causal links is based on Patel's tau measure (PT, Patel, Bowman, & Rilling, 2006, Smith et al., 2011). PT entails a two-step approach where the first step involves identifying the (undirected) connections by means of functional connectivity, and is achieved on the basis of correlations between the time series in different regions (which is also referred to as Patel's kappa, (Patel et al., 2006, Smith et al., 2011).

One note to make is that, Patel’s tau and other pairwise inference procedures assume that causation implies correlation. This assumption is necessary in order to perform the first step of the inference procedure, i.e. to select reliable connections for further classification into upstream and downstream regions. However, although this assumption is often true, this is not always the case as under certain circumstances (e.g. in control systems), causation might not be associated with correlation (Kennaway, 2009). In the recent study by Di & Biswal (2019), the authors investigated task modulated whole brain functional connectivity in six block-design and one event-related cognitive task. By using psychophysiological interaction between pairs of ROIs from the whole brain, the authors identified statistically significant task modulations in functional connectivity connectivity, and reported that 't*ask modulated connectivity was found not only between regions that were activated during the task but also regions that were not activated or deactivated*'. This suggests that only considering pairs of regions in which activity is correlated, might neglect some of the underlying effective connections.

There are multiple strategies for implementing and thresholding functional connectivity estimates (Varoquaux & Craddock, 2013). Since Pearson correlation typically returns dense connectomes that contain spurious links (spurious links X - Z appearing as a consequence of links X - Y and Y - Z, Aldrich, 1995), most often, partial correlation (Marrelec et al., 2006) is employed as a method of choice to build functional connectomes on the basis of fMRI datasets. In partial correlation, for each pair of nodes X and Y, the time linear input from all the remaining nodes in the network is *partialled out* from the time series before taking Pearson correlation. For sake of computational efficiency, partial correlation is often computed as inverse covariance.

Although partial correlation can be too conservative with respect to the underlying functional links or even induce spurious negative correlations in some cases (an effect known as Berkson's paradox (Nie, Yang, Matthews, Xu, & Guo, 2015, Berkson, 1946), to date, it remains the state of the art. Proper thresholding of partial correlation matrices is also an open research problem. A popular strategy for thresholding involves permutation testing (Smith et al., 2011, Hyvärinen & Smith, 2013), in which subject labels are shuffled between samples in multiple random iterations, and a separate null distribution is created for each connection in the network on the basis of these permutations (leading to a single thresholded connectome matrix for the entire subject population). Recently, also a thresholding strategy employing mixture modelling has been proposed (Bielczyk et al., 2018). These allow for individualized sparse connectomes to be derived per subject. Other alternative thresholding schemes involve proportional thresholding (van den Heuvel et al., 2017) and shrinkage estimates, such as Ledoit-Wolf (Ledoit & Wolf, 2004) and graphical Lasso (Friedman, Hastie, & Tibshirani, 2008).

Thresholding a functional connectome obtained with use of partial correlation results in a binary graph of connections, where all other possible edges identified as absent are disregarded from further considerations. The second step determines the directionality in each one of the previously detected connections. In this step, effective connectivity boils down to a two-node Bayesian network. The concept is based on a simple observation: if there is a causal link X -> Y, Y should get a transient boost of activity every time X increases activity. Vice versa: if there is a link Y -> X, X should react to the activation in Y. Therefore, one can threshold the signals X(t), Y(t) in order to obtain binary series of events $X_1(t)$, $Y_1(t)$, and compute the difference between conditional probabilities $P(Y_1|X_1)$ and $P(X_1|Y_1)$. Three scenarios are possible:

1. $P(Y_1|X_1)$ equals $P(X_1|Y_1)$: it is a bidirectional connection X <-> Y (since empty connections were sorted out in the previous step)
2. the difference between $P(Y_1|X_1)$ and $P(X_1|Y_1)$ is positive: the connection X -> Y is likely
3. the difference between $P(Y_1|X_1)$ and $P(X_1|Y_1)$ is negative: the connection Y -> X is likely

More recently, the Pairwise Likelihood Ratios (PW-LR, Hyvärinen & Smith, 2013) approach was proposed. PW-LR builds on the concept of PT. The authors improved on the second step of the inference by analytically deriving a classifier to distinguish between two models X -> Y and Y -> X, which corresponds to the LiNGAM model (Shimizu, Hoyer, Hyvärinen, & Kerminen, 2006) for two variables. The authors compare the likelihood of these two competitive models derived under LiNGAM's assumptions (Hyvärinen, Zhang, Shimizu, & Hoyer, 2010), and provide a cumulant based approximation to the likelihood ratio. Within the PW-LR family, there are three methods based on a third cumulant built for two signals X(t) and Y(t):

1. 'PW-LR skew' which is a classic third cumulant

2. 'PW-LR tanh' building an alternative approximation to third cumulant using a tanh-based non-linear correlation
3. 'PW-LR r skew' which contains an additional term discounting the outliers
4. 'PW-LR kurt' based on the fourth cumulant

The PW-LR approach clearly outperforms all the previously tested methods on the synthetic benchmark datasets (Hyvärinen and Smith, 2013). However, each one of the PW-LR methods is based on a single higher-order cumulant based on the BOLD distributions in two signals X(t) and Y (t) (either third or fourth order), it is not robust as lower moments can also account for possible differences in local signal-to-noise ratio (SNR). Furthermore, the SNR magnitude can differ with respect to the various features of the underlying neuronal time series, and if this is the case, these methods can erroneously flip the direction of the connection.

There are also other statistical methods for pairwise causal inference on the basis of a combination of marginal and conditional probabilities such as information-geometric approach by Janzing et al. (Janzing et al., 2012) or Unsupervised Inverse Regression by Sgouritsa et al. (Sgouritsa, Janzing, Hennig, & Schölkopf, 2015). However, we will not compare our method directly to these methods, as they have not been applied to fMRI datasets to date.(downstream) node. We demonstrate that our classifier gives higher robustness against these confounds than other methods.

Therefore, we further expand on the concepts of PT and PW-LR by proposing a classifier based on complex cumulants derived from multiple, possibly fractional, moments of the distribution of BOLD recordings. We compare the performance of our approach on synthetic benchmark datasets (Smith et al., 2011) relative to other methods for effective connectivity used in fMRI research. Furthermore, we compare performance of the methods using simple two-node simulations generated from the DCM model with varying signal magnitudes and noise variance in the projecting (upstream) and the receiving (downstream) node.

One needs to make a note about bidirectionally of connections here. As introduced above, Patel's tau approach (Patel, Bowman, & Rilling, 2006) involves a two step procedure. In the first step of this analysis, functional connectivity by means of partial correlation is used to find position of connections in the connectome (regardless of their directionality). The idea is, once a connection is found by means of functional connectivity, we can assume it can either be uni- or bidirectional. In the second step of the analysis, the directionality of a connection is determined through classification of two nodes into an 'upstream' and a 'downstream' node. Then, under the assumption that the tested method can correctly estimate the directionality of a connection, if the method does not give a univocal answer, we can assume that the connection is bidirectional (as in the first step of the analysis, connection was already detected). However, in case there are reciprocal connections between the two nodes, but one of them is stronger than the other, Patel's tau will only detect the 'net' effect, namely, it will indicate the stronger between the two connections. In that sense, reciprocal connection will be reduced to a univocal connection. The same effect holds for other pairwise inference methods, namely Pairwise Likelihood Ratios (PW-LR, Hyvärinen & Smith, 2013) and for our approach. In fact, reciprocal connections are ubiquitous in neuronal networks (Kötter & Stephan, 2003). Therefore, proper representation of these bidirectional connections remains an important challenge in network neuroscience.

We demonstrate that our classifier shows higher robustness against these confounds than other methods. In section 2.1, we introduce the concept of fractional moments of the distribution, and in section 2.2, we report on the procedure of combining fractional moments into fractional cumulants. In section 2.3, we give details on the classifier built on the basis of a set of fractional cumulants, and in section 2.4, we describe the supervised learning procedure. In section 2.5, we list popular methods for effective connectivity research in fMRI used for comparison. Finally, in section 2.6, we describe the generation of additional synthetic data with confounds often encountered in fMRI datasets, which we further use to benchmark the methods.

In section 3.1, we describe the results of the validation of our method using synthetic benchmark datasets (Smith et al., 2011). In Supplementary material 5, we present the detailed results of this validation. Furthermore, in section 3.2, we describe the results of an additional validation performed with use of the DCM generative model, but in presence of confounds such as a background noise and variability in hemodynamic responses between upstream and downstream region. Finally, in section 4, we critically discuss the results.

## 2 Materials and Methods

### 2.1 Fractional moment series of BOLD signal distributions

In this work, we propose to estimate causal links from BOLD recordings by analyzing the dependence between an expanded set of (fractional) moments of the BOLD distribution. We keep the same two-stage scheme for the causal discovery as proposed by Patel (Patel et al., 2006) and implemented by (Hyvärinen & Smith, 2013). However, we improve on the second step of the causal inference - a two-node classification problem - by utilizing all moments of the BOLD fMRI distributions, and combining them into cumulants.

Fig. 1 presents a graphical representation of a dynamical system with just two nodes. In this problem, one region ('upstream') is sending information to another region ('downstream') through a connection of weight $A_{12}$. Both regions receive region-specific signal $u_i(t)$. This signal can both relate to experimental input, as well as input from other regions. Both nodes also are influenced by a background neuronal noise $\sigma_i(t)$ that influence local signal-to-noise ratios.

The BOLD dynamics of such a simple two-node networks can be simulated with the DCM generative forward model (Friston et al., 2003). Given the inputs to the network, the connection strength and sets of parameters characterizing local hemodynamic response within the nodes, the DCM generative model makes prediction on the BOLD dynamics in both nodes. The details of the model are provided in Supplementary material 1. The ensuing BOLD time series can then be normalized, and characterized in terms of its central moments:

$$M_k = \frac{1}{N}\sum_{i=1}^{N} \hat{x}_i^k \quad (1)$$

where k in $Q$, $\hat{x}_i^k$ - time series with the mean and variance normalized to 0 and 1, respectively, N - the length of BOLD time series. The novelty in this approach is that a discrete set of moment orders k typically used to characterize a distribution (in terms of mean, variance, skew etc), is converted into a (pseudo-) continuous dimension, by sampling moment order k from sets of rational numbers. These fractional moments of the distribution are not isomorphic with the moment generating function [9] as we do not convert the distribution of BOLD values into a probability density function at any stage.

Since the original BOLD time series is normalized to mean of 0 (and variance of 1), it contains negative values and therefore, the fractional moments will become complex. Since the Eq. 1 is continuous with respect to k, these fractional moments will form a curve in the complex plane (Fig. 1 B).

In Fig. 1 B, we present a phase diagram for all moments in range k in [0.0, 5.0], for a long simulated BOLD fMRI representing a simple two-node network in the noiseless case. The moments are computed separately for the upstream region (blue) and the downstream region (red). The curve starts at (1,0) for k = 0. Then, it traverses the upper half-plane and for demeaned data arrives at (0,0) for k = 1. Subsequently, it goes back through the lower half-plane and comes back to (1,0) for k = 2 since the variance is equal to 1. Every time k becomes an integer, the curve crosses the real axis. Note that the imaginary axis characterizes the left half of the BOLD distribution, since fractional moments give nonzero imaginary part for negative values of distribution of the BOLD values.

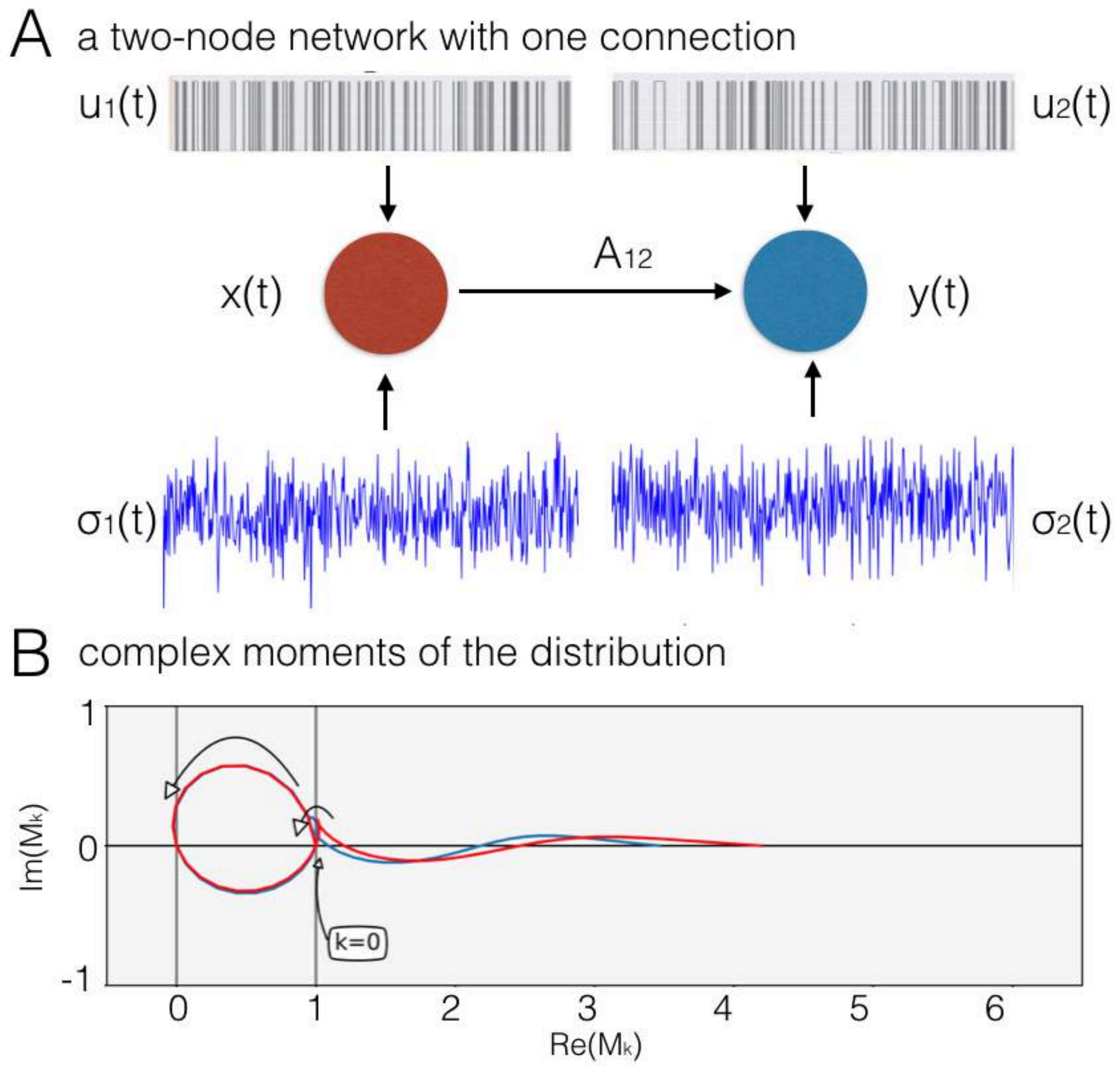

**Figure 1** A two node network with one directed connection. A: The upstream node, x(t), is sending information to the downstream node, y(t), through a single connection of weight $A_{12}$. Both regions received a binary signal $u_i(t)$ and neuronal noise $\sigma_i(t)$. The proportion between the amplitude of $s_i(t)$ and the variance of the noise $\sigma_i(t)$ defines the SNR in the network. B: All the fractional moments for k in [0; 5], for the BOLD fMRI time series from a simulated 2-node network, in the noiseless case. Blue: upstream region. Red: downstream region. The curve starts from (1, 0) for k = 0, traverses the upper half-plane and arrives at (0, 0) for k = 1. It travels back through the lower half-plane towards (1, 0) for k = 2 as the BOLD variance was fixed to 1 through the normalization. Every time k becomes an integer, the curve crosses the real axis.

## 2.2 Complex cumulants of the distribution

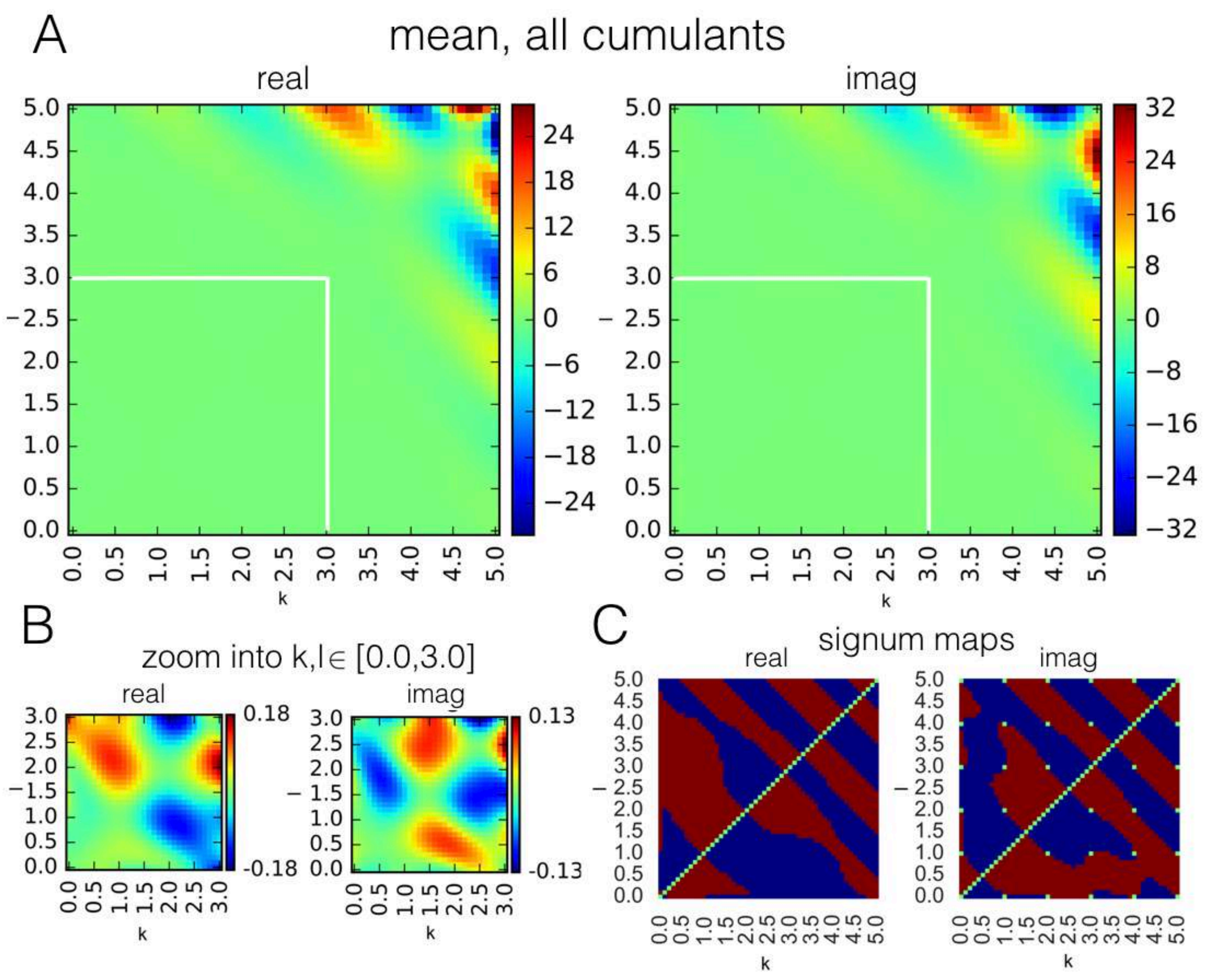

**Figure 2** **A** mean values for all cumulants, over 1000 simulations of a two node network with one connection (Fig. 1), for k, l in [0.0, 5.0]. Since cumulants are antisymmetric with respect to indexing k, l, the heatmaps for the real and the imaginary component are also antisymmetric. **B** a zoom of A into a smaller range of [0.0, 3.0]. The imaginary component is equal to zero for cumulants of integer orders: k, l in N, but exchanges the sign in the intervals between in a systematic way. C: sign of the cumulants in the majority of 1000 instantiations of the generative model. Red: positive. Blue: negative. Green: zero. We further denote the 'signum maps' for real and imaginary components as *Sr* and *Si*.

For two time series x(t), and y(t), not only the sole fractional moments but also the asymmetry between the moments can indicate the directionality of a connection. This asymmetry can be quantified by 'fractional cumulants':

$$C_{kl} = \frac{1}{N} \sum_{i=1}^{N} (\hat{x}_i^k \hat{y}_i^l - \hat{x}_i^l \hat{y}_i^k) \quad (2)$$

where k, l in $Q$.

In this particular problem, x(t), y(t) denote the BOLD time series in the two-node system. In order to make a prediction ('upstream' versus 'downstream') we learn the dependencies between moment time series using simulations on the basis of the DCM generative model. We have run 1,000 2-node DCM simulations with Fs = 200 [Hz] for a duration of 10 [min]. In order to marginalize out the influence of the hemodynamic parameters from our results, we sampled the parameters independently for the two nodes, and from the empirical distributions (Friston et al., 2003). In order to marginalize out the effect of different input strengths and frequencies, we also sampled the input magnitudes and frequencies (probabilities of switch from on- to off - state and vice versa) from a Gamma distribution with mean and variance of 1.

The input signals driving the upstream and the downstream region were also sampled independently from each other, as trains on- and off- states governed by Poissonian processes. The background neuronal noise was set to 0 in these simulations. In order to obtain a precise estimation of fractional cumulants, we did not sub-sample our synthetic BOLD every 2-3 [s] as is typically done for the synthetic BOLD fMRI.

We performed this simulation twofold. Firstly, we fed in an empty connection in order to create a null distribution of cumulants. Secondly, we added a connection $A_{12}$ with a weight of 0.9 to the 2-node system. In Fig. 2, we demonstrate the mean values for all cumulants of indexes k, l in [0.0, 5.0] obtained from the simulations of a connection, in Cartesian coordinates (we also performed similar computation in polar coordinates, however it did not give a substantial improvement to the classification performance, the results are presented in the Supplementary Material 3).

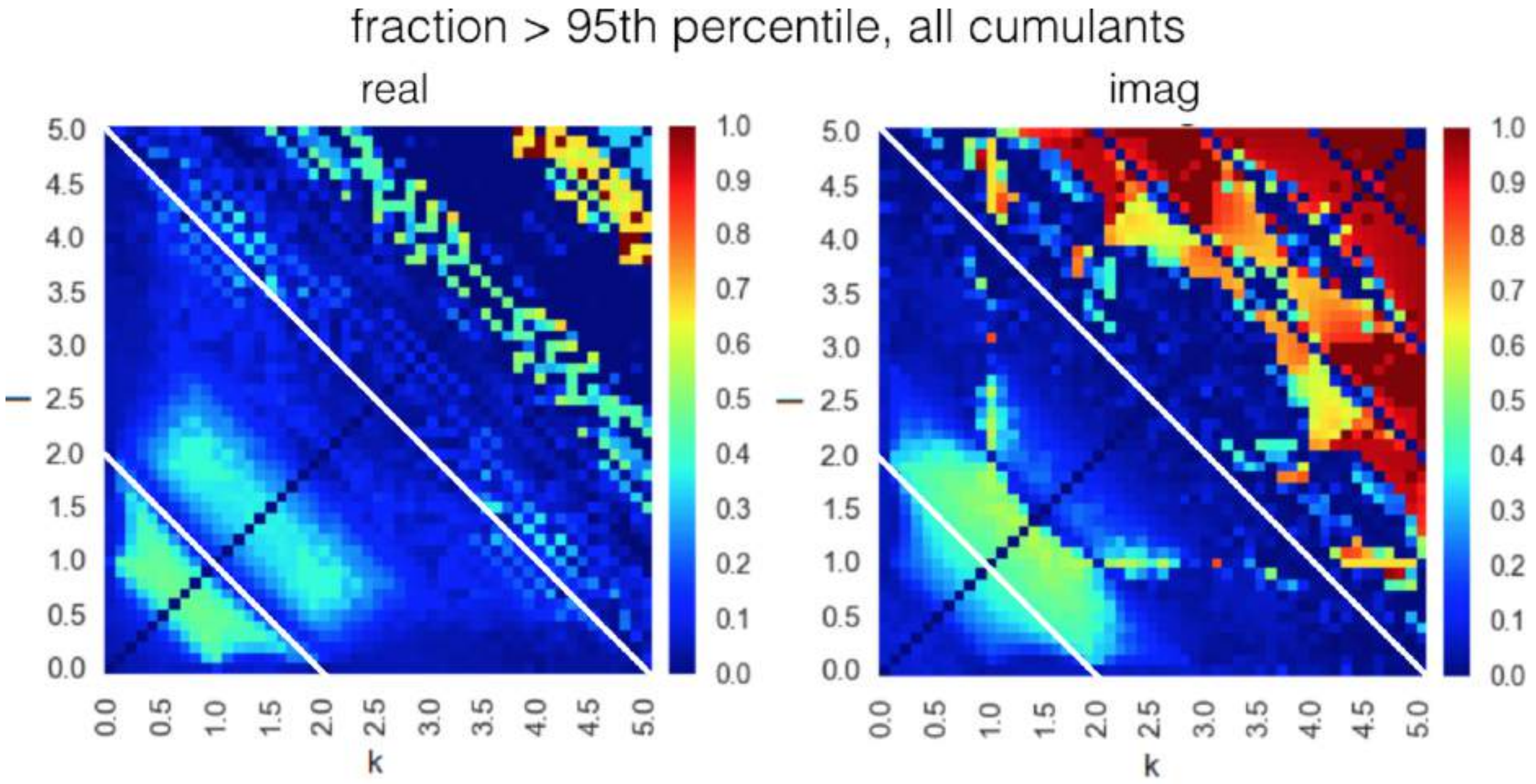

**Figure 3** Discriminative power for all cumulants in range k, l in [0.0, 5.0], in the ideal case of a very long BOLD time series and no background neuronal noise. Maximal discriminability in both real and imaginary cumulants exists in the range of k + l > 5.0. Additionally, in real cumulants there is a region of high discriminability in the range k + l < 2.0, and in imaginary cumulants there is such region in the range 2.0 < k + l < 5.0 (both boundary lines marked with a white line).

Fig. 2 A shows the mean values for all cumulants in range k; l in [0.0, 5.0], over 1000 simulations of a two node network with one connection (Fig. 1). Fig. 2 B shows the same maps are zoomed into the range k,l in [0.0, 3.0]. Finally, for every cumulant, Fig. 2 C, presents the sign of this cumulant for the majority of 1000 instantiations of the simulation (we further refer to these binary maps as Sr and Si).

These binary maps do not represent confidence intervals, or discriminability, for a particular cumulant, as 'majority' could mean 51% as well as 100% simulations. In order to choose cumulants which can best discriminate between a 'connection' and 'no connection' case, we created distributions of cumulant values across the 1,000 simulations in the null case and compared against the dis-tributions derived from simulations with a non-zero connection. We smoothed these distributions with kernel smoothing function and, for each cumulant, we computed the percentile of samples falling beyond 95th percentile of the null distribution (in case the mean for the given cumulant is negative as in Fig. 2 C, we took samples falling lower than the bottom 5th percentile of the null distribution, and higher than the 95th percentile otherwise).

The results of the discriminability analysis in Cartesian coordinates are shown in Fig. 3 (the results for polar coordinates are presented in Supplementary material 3). We can observe that whenever one of the indexes k, l equal to zero, i.e., the cumulant reduces to a simple moment, it has lower discriminative value than the full cumulants. Therefore, we will disregard moments from further analysis and fully concentrate on full cumulants, i.e., asymmetry between moments ($k, l > 0$).

Interestingly, the range of high discriminability is different for real and imaginary components. Maximal discriminability in both real and imaginary cumulants exists in the range of $k + l > 5.0$. Additionally, in real cumulants there is a region of high discriminability in the range $k + l < 2.0$, and in imaginary cumulants there is such region in the range $2.0 < k + l < 5.0$ (Fig. 3, both boundary lines marked with a white line). This different characteristic along the imaginary axis illustrates that moving from integer to fractional moments of the distribution provides with additional predictive power. Note also that, as we are deriving the discriminability maps from the DCM generative model, these maps are fingerprints of the particular problem of effective connectivity in fMRI; When derived from another generative model simulating another dataset, the maps would be different.

In order to investigate how the performance of classification based on single cumulants changes when a single connection is embed in a bigger network, we evaluated their success rate in estimating connectivity for benchmark synthetic datasets (Smith et al., 2011). Fig. 4 presents the grand mean success rate achieved by every cumulant separately, across all 28 benchmark synthetic datasets (Smith et al., 2011).

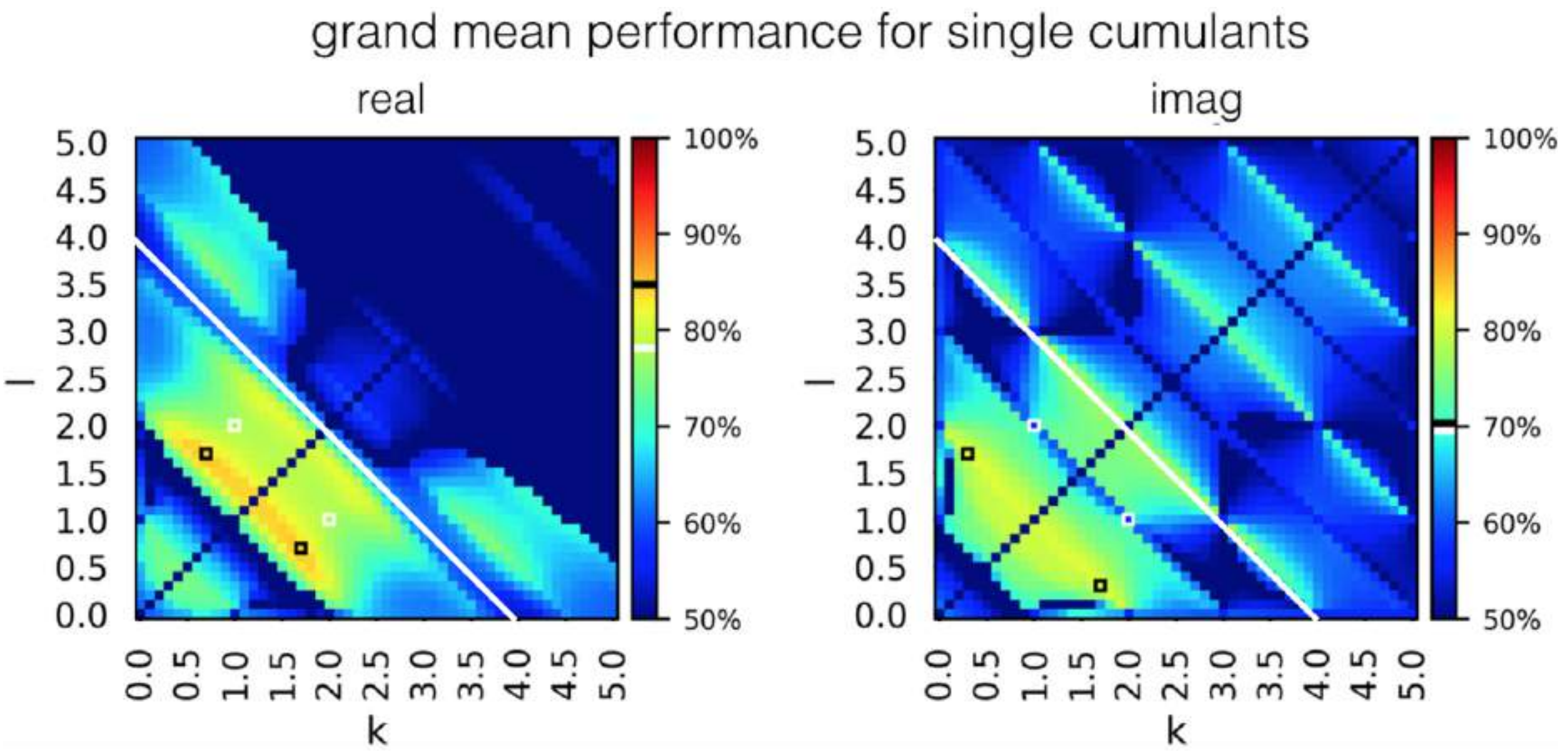

**Figure 4** Success rate for all the individual cumulants, averaged over 28 simulations from the synthetic benchmark datasets (Smith et al., 2011). White-edged square denotes a single cumulant used by Hyvärinen and Smith (Hyvärinen & Smith, 2013). The performance of this cumulant is shown in the colorbar, white band. Black-edged squares denote cumulants which give the highest performance on this dataset. Their performance is presented in the colorbar, black band. For the cumulants of high indexes k + l > 4.0 (white line), the success rate is not as high as the discriminability presented in Fig. 3 would suggest. The high success rate is not preserved for high-indexed cumulants which achieved high discriminability on 2-node simulations. The maximal grand mean performance equals 0.847 for the real components, and 0.814 for imaginary components.

The success rate for each of the 28 separate synthetic datasets (Smith et al., 2011) is presented in the Supplementary material 4. The maps of simulation-dependent success rate relate to the maps of discriminative power (Fig. 3), but they are not identical and differ between simulations. One difference is that for the cumulants of high indexes k + l > 4.0, the success rate is not as high as the discriminability presented in Fig. 3 would suggest. This is because Fig. 3 represents the limit of a system of two isolated nodes with infinite SNR, and a very long BOLD time series, whereas bench-mark synthetic datasets refer to more realistic case when for each pair of nodes, the time series is short, there are confounding signals from other nodes in the network, and there is certain degree of noise in the communication (see: Supplementary material 2). Altogether, these factors cause that the high moments are hard to estimate in practice.

### 2.3 Combining fractional cumulants into a classifier

We propose to combine information contained in multiple cumulants by building the classifier based on a 'voting' scheme between the cumulants. This classifier determines whether the map of cumulants obtained for a pair of time series X(t), Y(t) is closer to the benchmark maps presented in Fig. 2 A (which is an evidence for a connection X -> Y), or their inverse (which is an evidence for a flipped connection Y -> X). Each of the cumulants $C_{kl}$ votes due to sign $Sr_{k,l}$, $Si_{k,l}$ (Fig. 2 C). If the sign of the cumulant is the same as in Fig. 2 C, it adds to the evidence for a connection X -> Y, and against this connection otherwise.

Since in realistic conditions (short datasets, large TRs), high index cumulants, k + l > 3.0, yield the aforementioned estimation problem, we discount their contribution in the voting by using a non-linearity of a form (further referred to as *weighting* throughout the manuscript):

$$f(x)=\log(\cosh(\max(x,0))) \quad (3)$$

Similar function was proposed to discount the outliers present in the BOLD time series in the work by Hyvärinen & Smith, 2013. The final classifier yields:

$$\begin{cases} X \rightarrow Y & \text{if} \sum_{k,l} [\mathrm{Sr}_{k,l} \mathrm{f}(\mathrm{Cr}_{k,l}) + \mathrm{Si}_{k,l} \mathrm{f}(\mathrm{Ci}_{k,l})] \geq 0 \\ Y \rightarrow X & \text{otherwise} \end{cases} \quad (4)$$

in the weighted case, and

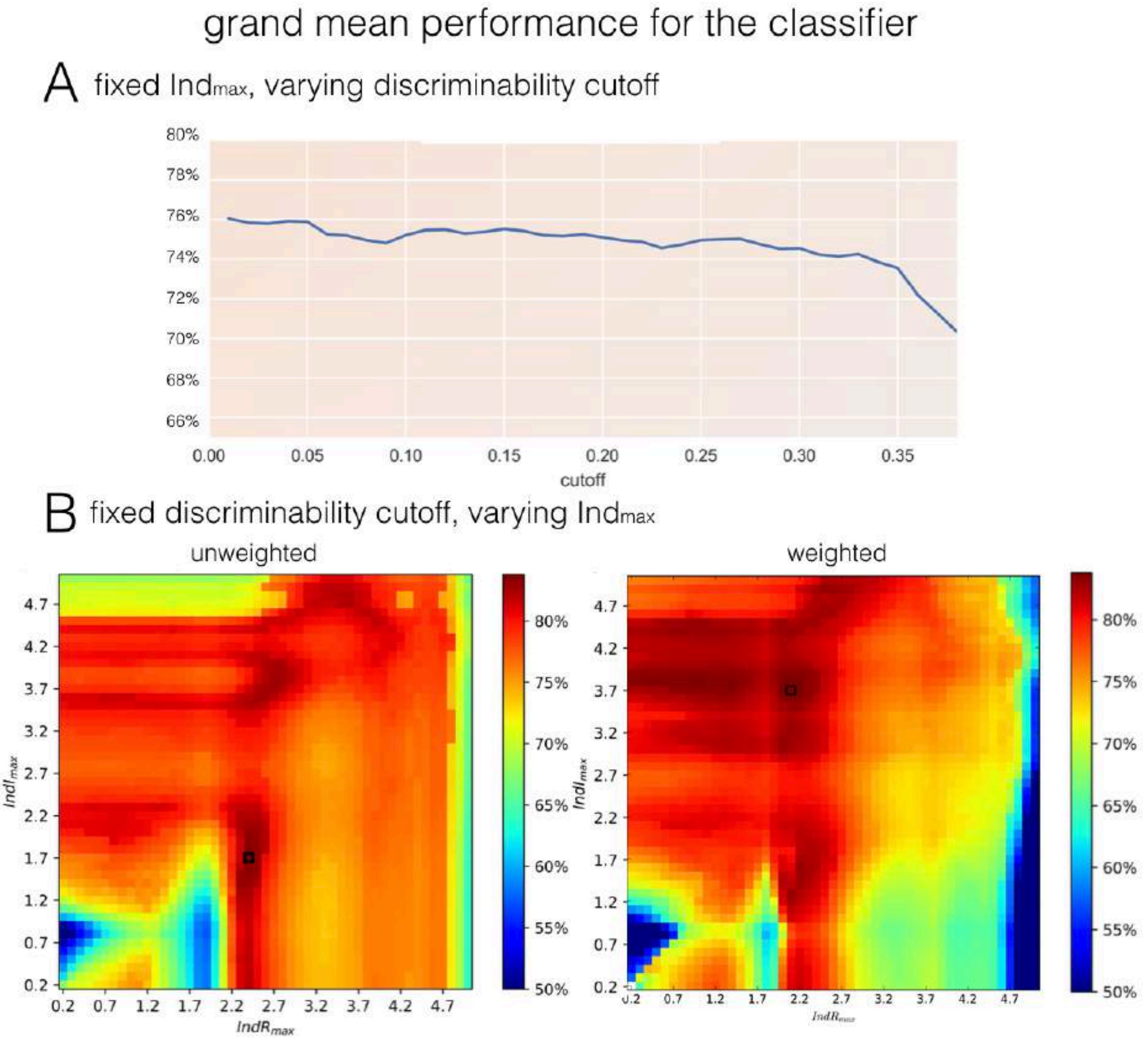

**Figure 5** Dependence of the grand mean performance on synthetic datasets on the choice of cumulants. A: Grand mean performance for unweighted cumulants in range k + l <= 3.1, in the function of the cutoff discriminative value. The higher cutoff, the less cumulants we take into account while voting for the directionality of the link. The results clearly show that in order to maximize the success rate in estimating effective connectivity, we should all the cumulants should be taken into account, except for the diagonal of k = l. B: the grand mean performance based on cumulants of indexes k, l between 0.1 and k + l <= $Ind_{max}$, in the function of that maximal index. The optimal performance in unweighted case equals 0.835 for [$IndR_{max}$; $IndI_{max}$] = (2.4, 1.7), and 0.886 for [$IndR_{max}$; $IndI_{max}$] = (2.1, 3.7) in weighted case.

$$\begin{cases} X \rightarrow Y & \text{if} \sum_{k,l} [\mathrm{Sr}_{k,l}\mathrm{Cr}_{k,l} + \mathrm{Si}_{k,l}\mathrm{Ci}_{k,l}] \geq 0 \\ Y \rightarrow X & \text{otherwise} \end{cases} \quad (5)$$

in the unweighted case.

**2.4 Supervised learning using synthetic benchmark datasets**

In this work, we derive the classifier using sign maps $\mathrm{Sr}_{k,l}$, $\mathrm{Si}_{k,l}$ (Fig. 2 C) developed using multiple realisations of a 2-node simulation under the Dynamic Causal Modeling generative model, which is a form of a supervised learning. In a different application and under a different generative model, these maps would look differently.

We know that cumulants differ with respect to discriminability (Fig. 3), and that the success rate of cumulants differs depending on the range $k + l <= \mathrm{Ind}_{max}$ (Fig. 4). Therefore, we optimize the performance of the classifier with respect to these two dimensions, by finding a combination which gives the best grand mean performance across the 28 simulations from the synthetic benchmark datasets, as they represent the variety of experimental conditions encountered in real-life fMRI setups.

Firstly, we fix $\mathrm{Ind}_{max} = \max(k, l) = 3.1$, and consider cumulants on a triangle $k; l >= 0.1$, $k + l < \mathrm{Ind}_{max}$. We then choose only cumulants of discriminability exceeding a particular value to be fed into the classifier. For instance, cutoff value of 0.1 means that we include the vote from all cumulants for which the discriminative value is not less than 0.1. We can then evaluate the grand mean success rate (as the mean success rate over all 28 benchmark synthetic datasets) in the function of the thresholding discriminability value. Fig. 5 A, demonstrates that including all the cumulants with a positive discriminative value (all cumulants except for $k = l$, for which discriminability is always zero) gives the best classification performance.

Secondly, we optimize the window $\mathrm{Ind}_{max}$ for indexes k, l and compare between classifier with and without weighting with the discount function introduced in Eq. 3. Since discriminability is generally higher for low indexes k, l (Fig. 3), we will evaluate the grand mean performance based on cumulants of indexes between 0 and a maximum $Ind_{max}$, in the function of that maximum. We consider the maximal indexes along real and imaginary dimension separately. The results are presented in Fig. 5 B. The optimal performance in unweighted case equals 0.835 for [$IndR_{max}$, $IndI_{max}$] = (2.4, 1.7), and 0.886 for [$IndR_{max}$, $IndI_{max}$] = (2.1, 3.7) in weighted case, which exceeds both the grand mean performance of the 'PW-LR skew r' method by Hyvärinen (0.845) and the maximal grand mean performance of any single cumulant in our study (Fig. 4, the maximum of 0.847).

**2.5 Selection of other approaches for effective connectivity research in fMRI**

In order to benchmark our classifier, we compare the performance against other methods. As mentioned in the Introduction, the field of effective connectivity in fMRI is very wide (Smith et al., 2011), therefore, we restricted ourselves to the most popular approaches (other than DCM itself, Friston et al., 2003), as the synthetic datasets are generated straight from the DCM generative model):

1. state-space implementation of Granger Causality (GC, Granger, 1969, Seth, Barrett, & Barnett, 2015), which is a multivariate method inferring effective connectivity between a pair of time series under the assumption that both of them can be expressed as autoregressive processes. We used simple version of Granger Causality featuring Ordinary Least Square regression with lag of 1, implemented in Multivariate Granger Causality Toolbox (Barnett & Seth, 2014), obtained from http://www.sussex.ac.uk/sackler/mvgc. For GC based on VAR process as in our study, the state-space implementation is more robust than spectral GC (Geweke, 1982, Geweke, 1984), because the frequency-domain version has a bias–variance trade-off (a function of the VAR model order that can induce spurious conditional GG causality estimates, such as erroneous peaks in the frequency domain, as indicated in the recent work by (Stokes & Purdon, 2017). Furthermore, the state space formulation of GC is the most robust, mitigating effects of bias and variance due to the fact that the reduced model is VAR rather than VARMA (Barnett & Seth, 2015). In order to compare performance with the methods for pairwise inference, we used GC in a bivariate rather than multivariate fashion: by applying GC to each of the previously found connections separately

2. Partial Directed Coherence (PDC, Baccalá & Sameshima, 2001). Partial Directed Coherence is known as a method conceptually close to Directed Transfer Function (DTF, Kamiński & Blinowska, 1991, Baccalá & Sameshima, 2001). Both these methods are derivatives from Geweke spectral measures of Granger causality (Geweke, 1982, Geweke, 1984) and all three methods have similar limitations (Chicharro, 2011). However, PDC is used substantially more often in fMRI studies than the other two methods, especially when compared to spectral Granger causality. For this reason, we chose PDC as a method representing this class of approaches. We used PDC implementation from the Extended Multivariate Autoregressive Modelling Toolbox (Faes, Erla, Porta, & Nollo, 2013): http://www.lucafaes.net/emvar.html. As in case of GC, we applied PDC in a bivariate fashion
3. Patel's tau (PT, Patel et al., 2006), as described in the Introduction. Patel's tau, implemented similarly as in Smith et al., 2011: by recalculating each time series into the range [0, 1], setting samples under the 10th percentile to 0, over the 90th percentile to 1, and linearly mapping the remaining samples to the range [0, 1]. Then, we infer the directionality of connection from the difference between P (X|Y) and P (Y|X). In addition to the previous implementation however, we also integrate the results over all the possible thresholds in order to eliminate the thresholding problem while calculating the conditional probabilities P (X|Y), P (Y|X)
4. Pairwise Likelihood Ratios methods (PW-LR, Hyvärinen & Smith, 2013) represented by 'PW-LR r skew', as it gives the highest performance among all the PW-LR methods when applied to synthetic fMRI data. We obtained the codes for the PW-LR methods from https://www.cs.helsinki.fi/u/ahyvarin/code/pwcausal/ (Hyvärinen used the cumulant k; l = (2; 1) weighted with covariance for synthetic benchmark datasets. Therefore, for this comparison, we use the classifier based on fractional cumulants weighted with covariance. For our study, we chose a 'PW-LR r skew' version of the method, which involves inference based on a third cumulant with a discount for outliers, Hyvärinen & Smith, 2013)

As in Hyvärinen & Smith, 2013, we performed the first step of the inference as inverse co-variance thresholded with permutation testing. All the methods, including multivariate methods such as GC and PDC, were then applied in a pairwise fashion (i.e., separately for each two-node node network representing a single connection found in the previous step).

Furthermore, we did not include the DCM procedure in this comparison, for the same reasons as Smith et al., 2011: DCM is not an exploratory method and is using it in this context, namely for exploratory causal research on the set of benchmark synthetic datasets (where the smallest network consists of 5 nodes) is not computationally feasible. Furthermore, the characteristics of this synthetic benchmark datasets is that, input signals (Fig. 1A) represent random events and can therefore emulate all types of fMRI experiments: classic task-fMRI studies, event-related responses or resting state BOLD time series. In DCM however, the inputs must be strictly specified in the block design, otherwise DCM inference cannot be initiated. Therefore, assumptions of DCM do not fit research problem formulated in this particular way.

**2.6 Testing robustness of the methods against confounds**

In addition to evaluating our approach against the existing simulations from Smith et al., 2011, we further evaluate the performance under additional yet typical modes of variation in the data. Specifically we are interested in characterizing the discriminative performance relative to (i) more complex forms of stochastic noise in the data and (ii) unequal levels of SNR per node.

The benchmark synthetic datasets involve temporally uncorrelated, white background noise of a low magnitude on the neuronal level (Smith et al., 2011). This type of noise is not physiologically plausible, as it is known from physiological studies that in the neuronal networks, the back-ground noise has a scale-free power spectrum (He, 2014, Bédard et al., 2006, Dehghani et al., 2010, Bielczyk et al., 2017). Therefore, we simulated a two node system and introduced scale-free (pink) noise to the system. Then, we varied the variance of the noise in the range of [0.2, 5.0] while keeping the amplitude of the inputs $s_i(t)$ fixed to 1.0. We performed 500 realizations of 10[min] simulation at high temporal resolution of Fs = 200[Hz], for each configuration of the noise variances.

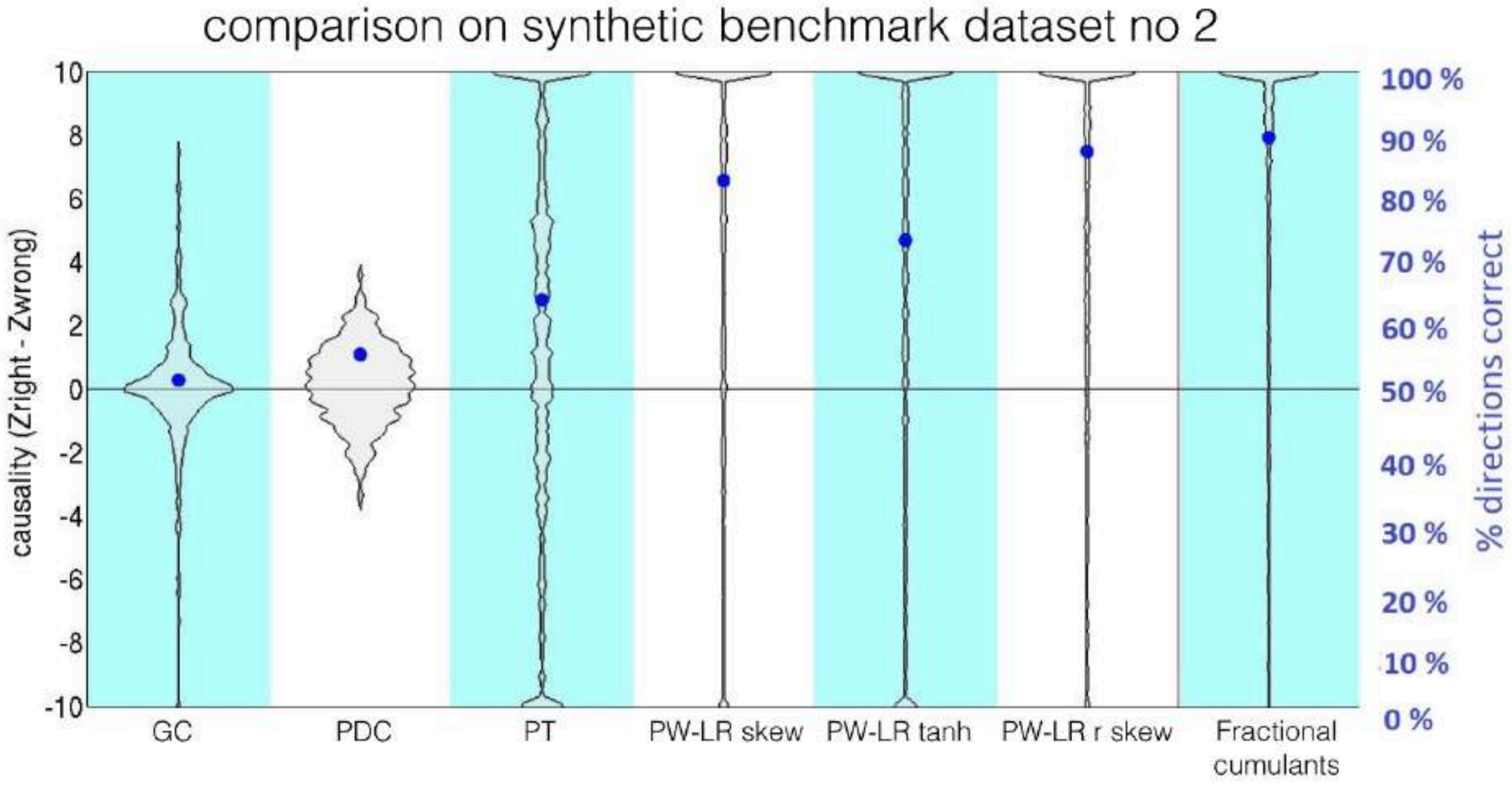

**Figure 6** Comparison between the classifier based on the fractional cumulants and several other methods, on benchmark simulation no 2. The violin plots denote the distribution of the z-scores for connections as compared to the null distribution. Blue dots denote the percentage of correct assignments for the true connections (Smith et al., 2011). The difference in performance between the classifier based on fractional cumulants and 'PW-LR r skew' (Hyvärinen & Smith, 2013) is small.

Furthermore, in the original version of the DCM procedure (Friston et al., 2003), as well as in most computational studies (Smith et al., 2011), equal stimulus strengths to both nodes $s_i(t)$ are assumed. This assumption might not hold true in the real fMRI datasets. Therefore, we performed another, noiseless simulation, in which we varied signal strengths between the upstream and down-stream region. We performed 500 realizations of 10[min] simulation at high temporal resolution of Fs = 200[Hz], for each configuration of input strengths in the range of [0.2, 5.0].

## 3 Results

### 3.1 Supervised learning using synthetic benchmark datasets

The best version of the classifier was obtained for voting between cumulants in the range [$IndR_{max}$; $IndI_{max}$] = (2.1; 3.7), with the discount for high moment indexes (Eq. 3). The comparison of this classifier against four other methods, GC, PDC, PT and 'PW-LR r skew', on the benchmark simulation no 2 is presented in Fig. 6. The violin plots denote the distribution of the z-scores for connections as compared to the null distribution. Blue dots denote the percentage of correct assignments for the true connections, as in (Smith et al., 2011). In most of the other 27 benchmark datasets, our classifier outperforms all the other methods (Supplementary material 5). As in the original study by Smith et al., 2011, lagged methods, GC and PDC, perform worse than the structural methods. 'PW-LR r skew' and fractional cumulants both outperform PT, most probably because PT is based on the thresholded signal and therefore contains a free parameter.

In general, in the benchmark synthetic datasets, fractional cumulants outperform all other techniques in almost all cases, although the difference between performance of the fractional cumulants and PW-LR methods is small.

### 3.2 Robustness of the methods with respect to confounds

Fig. 7 presents the comparison between fractional cumulant classifier and various other methods on a 2-node simulation under the addition of varying levels of physiologically plausible, scale-free neuronal noise across various levels of SNR for the upstream and downstream region. The results suggest that all previously tested methods show low levels of robustness towards such additional sources of variability in the data. GC as well as PDC are at the lowest performance, and give results on a chance level across the whole parameter space. In PT, the success rate in proper classification between upstream and downstream node rapidly drops towards 50% along with a decrease in SNR in the upstream node. However, in case SNR in the upstream node is higher than 1 (left half of the 'Patel's tau' panel, Fig. 7), PT is resilient to the variance of SNR in the downstream node. The performance of 'PW-LR r skew' drops down to the chance level with respect to both the noise level in the upstream and downstream region, whereas GC and PDC are performing almost equally poorly under any combination of noise variances (probably because the variance of the noise from the fitted autoregressive model is used to establish the directionality of the causal influence in GC).

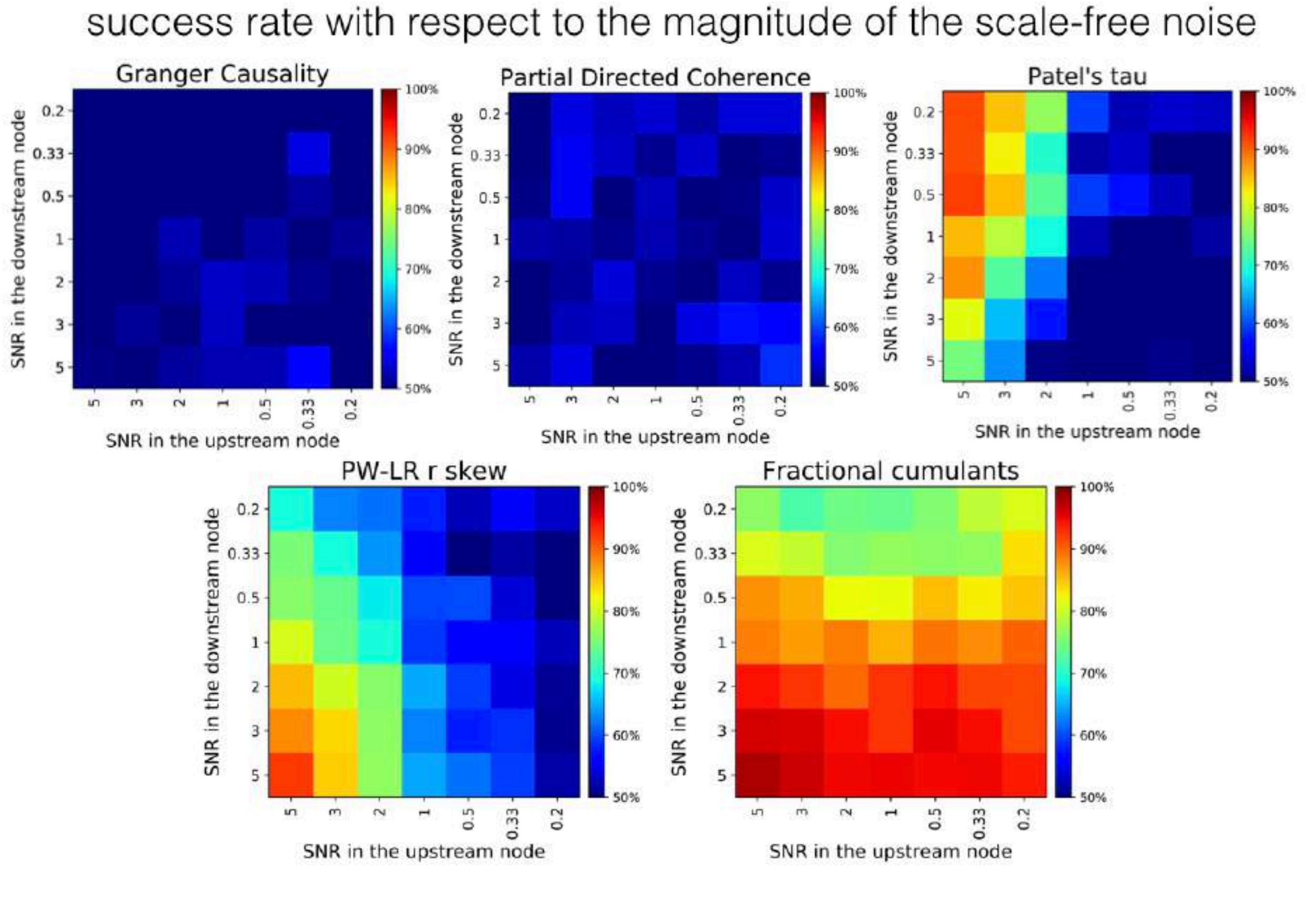

**Figure 7** Robustness of the methods against the background scale-free neuronal noise. The variance of the background noise σ(t) differs between upstream and downstream region in the range of [0:2; 5:0]. As signal magnitude is constant and equal to 1 in these simulations, the signal-to-noise ratio (SNR) was calculated as 1/var(σ).

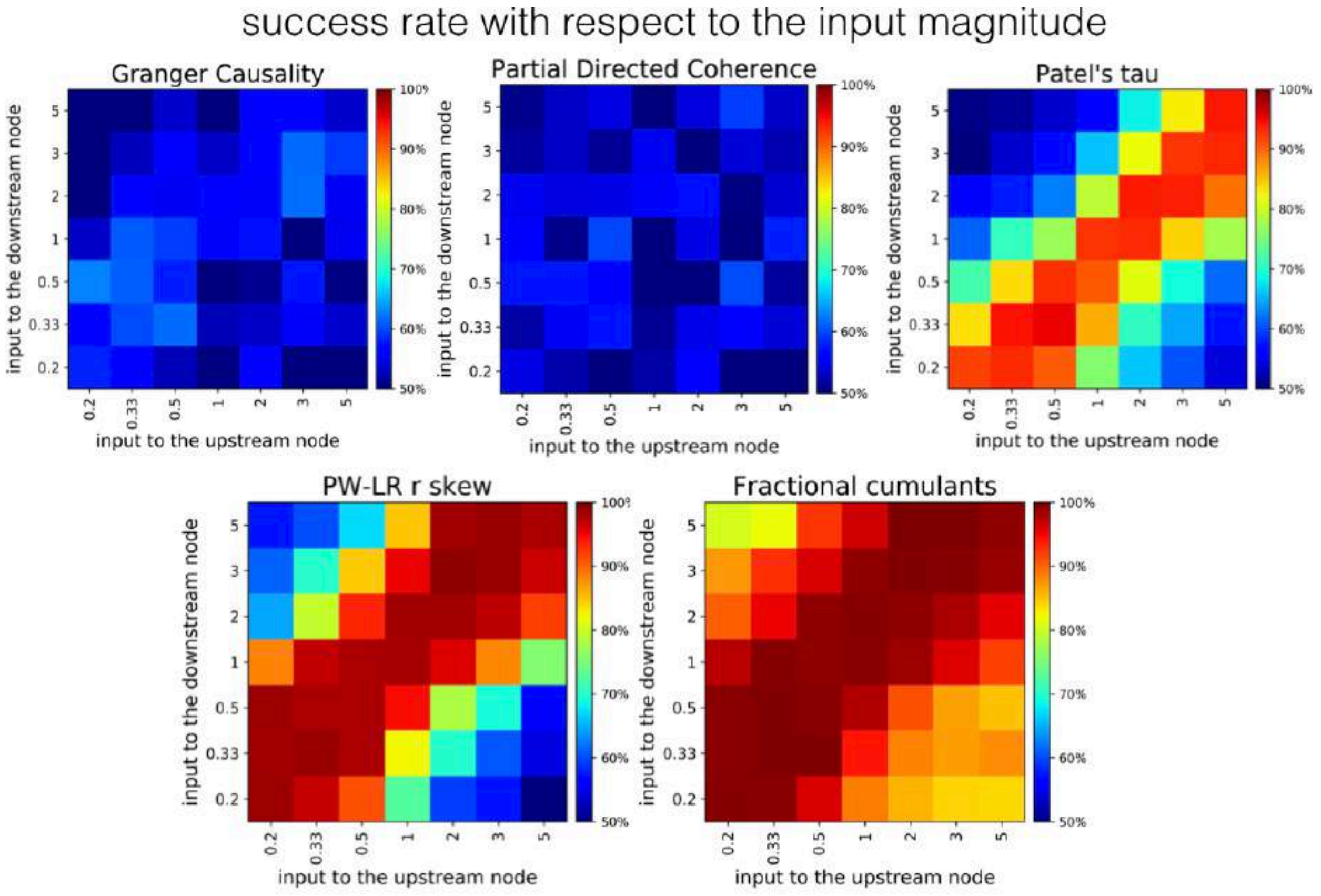

**Figure 8** Robustness of different methods to the change in signal strengths. The variance of the signal differs between upstream and downstream region, both in the range of [0:2, 5:0]. GC and PDC give performance around the chance level across the whole parameter space, whereas 'PW-LR r skew' and PT exhibit certain resilience towards this variability in the inputs. However, the classifier based on fractional cumulants is the only method whose performance does not fall towards the chance level within the parameter space.

Fig. 8 presents the comparison between the classifier based on fractional cumulants and other methods, given noiseless simulation and varying signal magnitudes. The classifier based on fractional cumulants is the only method where the performance does not decrease down to chance level within the chosen parameter space. GC and PDC give performance around the chance level across the whole parameter space, whereas 'PW-LR r skew' and PT exhibit certain resilience towards this variability in the inputs. However, the performance breaks down to the chance level in cases when signal magnitudes between the upstream and downstream node becomes highly disproportionate (higher than 3.0).

# 4 Discussion

## 4.1 Summary

This work provides an advance to the effective connectivity research in functional Magnetic Resonance Imaging, by utilizing the additional information contained in the BOLD time series with use of fractional moments of the BOLD distribution combined into cumulants. Usage of this additional information (embedded within a classifier) significantly increases the robustness towards plausible sources of variability in fMRI, namely presence of physiologically realistic (scale-free) background noise as well as disproportion in the inputs strength, either due to differences in the amount of neuronal activity locally induced and/or due to effective differences induced by e.g. regional variations in the coil sensitivity profiles. This is where the value of added information coming from fractional cumulants becomes apparent: among the methods tested in this work, only the classifier based on the fractional cumulants gives a performance better than chance across the whole parameter space.

Effective connectivity is a research problem directly related to the notion of causality. Causality is, in general, difficult both to define (Pearl, 2000) and to measure. In the most basic formulation, in X causes Y, it means that without X, Y would not occur. In practice, the picture is far less clear in complex dynamic systems such as brain networks: for any event, high number of potential causes can be defined, and these causes most often interfere with each other. This research problem was recently discussed by Albantakis, Marshall, Hoel, & Tononi, 2017 who decomposed causality into independent dimensions: realization, composition, information, integration and exclusion. Also, interpretation of a causal component in a given process depends on the context. In example, respiratory movement is typically considered a confound in fMRI experiments, unless we are interested in the influence of respiration speed on the activity of neuronal populations. Furthermore, in brain networks, temporal ordering of the cause of cause and effect is hard to maintain as information is circulating in recurrent rather than feed forward networks (Schurger & Uithol, 2015).

Furthermore, If we have an interventional study at disposal, establishing causality becomes straight forward, but this is rarely the case in human research. In human fMRI, all the studies are observational rather than interventional. In such case, causation can never be observed directly, just correlation (Hume, 1772). When a correlation is highly stable, we are inclined to infer a causal link. Additional information is then needed to assess the direction of the assumed causal link, as correlation indicates for association and not for causation (Altman & Krzywinski, 2015).

In the simulations, we used multiple realisations of the dynamics for every network pattern. We needed to run multiple instantiations of a noisy system in order to evaluate the mean success rate of the methods under noisy circumstances. We also simulated long runs of the dynamics aiming to find an *upper bound* on the methods' performance in a function of the SNR disproportion and background noise levels. We did not investigate the effects of the sample length on the results. This is because in our study, we focused on the confounding factors which - unlike the duration of the study - cannot be influenced by the researcher.

Although fractional moments of a distribution, as a mathematical concept, were studied before (Dremlin, 1994, Matsui & Pawlas, 2013), this concept was not applied to biomedical sciences to date. One reason for this lack of applications might be that the fractional moments become complex numbers for the normalized time series, and that subsequently, the features characterized by these moments cannot be conceptualized as easily as the features characterized by the integer moments (e.g., skewness can be interpreted as a measure of 'asymmetry' of the distribution, and kurtosis can be interpreted as its 'flatness'). However, although the fractional moments of a distribution are a mathematical concept with limited practical interpretation, but could still contain valuable (causal) information. In this work, we demonstrate that these features of the distribution provide important additional information about the distribution of BOLD values. We first perform supervised learning of the classifier on the set of benchmark synthetic datasets, and then validate the classifier on 2-node simulations with biologically realistic confounds. We believe that confounding factors such as a physiological background noise of a magnitude varying between the nodes is an important to overcome for any method for causal inference in fMRI. This is because every network in the brain is embedded in larger networks, therefore the background activity of other interconnected networks can be interpreted as 'noise' (Deco, Jirsa, & McIntosh, 2011). We demonstrate that our approach can increase the robustness of the methods for pairwise inference in fMRI to the main sources of variability in BOLD fMRI.

Unlike the previous methods for pairwise inference in fMRI (Hyvärinen & Smith, 2013), the classifier defined in this study is informed by the Dynamic Causal Modeling generative model, therefore it incorporates the priors derived from the neurophysiological studies (Buxton, Wong, & Frank, 1998). Deriving benchmark signum maps for the classifier from the multiple instantiations of the DCM generative model allows for marginalizing out all the parameters unimportant for the effective connectivity research: the classification procedure focuses only on classifying a pair of regions into upstream and downstream instead of fitting all the hyperparameters as is done in the classic DCM inference procedure. Therefore, this approach is a reduction of the problem of effective connectivity in a large network to a two-node classification problem on one hand, and an extension of the feature space from integer to fractional moments on the other hand.

The signup maps derived in the training process, are dependent on the generative model. In this study, we chose the canonical, original formulation of the Dynamic Causal Modeling (Friston et al., 2003) for the study. There are also newer formulations of the DCM, e.g. the canonical microcircuit approach (Pinotsis et al., 2017, Friston et al., 2017), in which layer-specific neuronal populations in the cortex are modelled with use of a neural mass model. In this case, we assume that the inference is performed on mesoscale level, in which ROIs represent brain regions (cortex regions or subcortical nuclei) rather than cortex components. Furthermore, although version of DCM containing higher-order, nonlinear effects (Stephan et al., 2008) are also developed, we believe that (bi)-linear model is a good simplification to describe the underlying neuronal dynamics, as it refers to the linear part of the sigmoidal transfer functions between neuronal populations in the brain (Silver, 2010, Bielczyk, Buitelaar, Glennon, & Tiesinga, 2015). Modeling communication between nodes in the network with use of linear transfer functions is a common practice in modelling effective connectivity in fMRI, see e.g. Structural Equation Models (McLntosh & Gonzalez−Lima, 1994) or LiNGAM-ICA (Shimizu et al., 2006, Smith et al., 2011). The bilinear version of DCM, often referred to as the general linear model approach, is still a very popular tool for finding effective connectivity patterns from fMRI in clinical practice (see, e.g., recent work by Zhang et al., 2018, Nackaerts et al., 2018, Arioli et al., 2018, Pool et al., 2018).

We reproduced the DCM generative model after (Smith et al., 2011). This implementation is acknowledged in the field as a benchmark setting for testing new methods for functional and effective connectivity in fMRI, see e.g. (Smith & Hyvarinen, 2013, Hinne et al., 2015, Bielczyk et al., 2018). In this implementation, bilinear effects (namely, modulation of connections by experimental inputs) are not modelled. In pairwise inference this omission is justified, as modulation of connection only affects the strength of connection weight $A_{12}$, which will influence cumulant values quantitatively and not qualitatively - which does not influence the outcome signup maps.

Evaluating methods with use of synthetic datasets as the ground truth is typically the first step in validation of any new data analytic framework. Validating new methods with use of synthetic datasets is a golden standard across the whole field of neuroimaging, from single cell imaging to whole-brain imaging with fMRI or EEG/MEG. This tradition has a long history, starting from the Nobel-winning Hodgkin and Huxley model for initiation and propagation of action potential (Hodgkin & Huxley, 1952). Today, methods for effective connectivity between neuronal assemblies measured with multielectrode arrays, are still validated on synthetic datasets generated from this classical model, including recent approaches: nonlinear data assimilation (Hamilton, Berry, Peixoto, & Sauer, 2013) and differential covariance (Lin, Das, Krishnan, Bazhenov, & Sejnowski, 2017). In cognitive neuroimaging, testing methods on synthetic datasets from generative models is also a standard. In EEG/MEG research, there are multiple classes of generative models generating different type of dynamics, depending on the purpose of the modeling study, e.g.: the nonlinear lumped-parameter model for generating alpha rhythms and its neural-mass extension by (David & Friston, 2003) Wong-Wang model for winner-take-all dynamics (Wong & Wang, 2006), Hindmarsh-Rose model for epileptor dynamics (Hindmarsh & Rose, 1984) and Dynamic Causal Modeling for EEG / MEG (Kiebel, Garrido, Moran, Chen, & Friston, 2009, Steen, Almgren, Razi, Friston, & Marinazzo, 2018, Moran, Pinotsis, & Friston, 2013). Furthermore, the Human Neocortical Neurosolver simulator developed at the Brown University (HNN, https://hnn.brown.edu), is a complex tool simulating Local Field Potentials measured with EEG / MEG by bottom-up modeling of clusters of neurons. All these tools can serve to validate new methods for functional and effective connectivity in EEG / MEG (Valdes-Sosa et al., 2011, Wang et al., 2014).

In fMRI, the selection of generative models is narrower than in EEG / MEG: the Dynamic Causal Modeling (DCM, Friston, Moran, & Seth, 2013, Smith et al., 2011) achieved a status of a standard generative model. With use of this synthetic data generated from this model, new methods for effective connectivity in fMRI are validated, e.g. the third and fourth cumulant method by Hyvarinen & Smith,  2013 and artificial immune algorithm combined with the Bayes net method (AIAEC, Ji, Liu, Liang, & Zhang, 2016).

To date, DCM is the most biologically relevant generative model proposed in the field of functional Magnetic Resonance Imaging. The implementation of benchmark synthetic datasets based on DCM by Smith et al., 2011 has gained a lot of attention and following in the field but also, it gained its critics. For instance, according to Smith's results, Patel's tau (Patel, Bowman, & Rilling, 2006) is one of the methods giving best performance in retrieving directed connectivity patterns from synthetic benchmark datasets. In a recent work, Wang, David, Hu, & Deshpande, 2017 performed modelling study on datasets derived from experiment by David et al., 2008)in which fMRI activity was measured in genetically modified rats suffering from epilepsy.

Activity from the same set of regions was recorded in an associated intracerebral electroencephalographic study, in order to provide with ground truth information flow. The authors chose primary somatosensory cortex barrel field (S1BF), thalamus and striatum (caudate-putamen; CPu) as ROIs and demonstrated that Patel's tau is no better than chance in recovering directional connectivity patterns from this data in both raw and deconvolved fMRI datasets. On the contrary, DCM and Granger Causality proved to correctly estimate the directionality of the information flow on the group level and on the deconvolved data.

There are more caveats with respect to the benchmark DCM simulations by Smith et al., 2011. Firstly, synthetic datasets derived by the authors of the study involve a low noise condition, in which the background noise in the networks in as low as 5% of the signal magnitude. Given that the background activity in the brain networks includes not only noise but also echo of cognitive processes unrelated to the experiment, 5% of background activity seems to be on the extremely low end of the spectrum of possibilities. Secondly, networks investigated by Smith et al. are sparse (a number of connections in a network of size N is of order of N) and almost acyclic, also seems to be a very optimistic scenario. Thirdly, Smith et al. used a TR of 3 [s] and time series of 200 data points. This TR is too long and the time series length too short for generalizing the empirical datasets used today. Today, shorter TRs (1 [s] or less, e.g. 0.72 [s] as in Human Connectome Project datasets, Van Essen et al., 2013) and longer time series have become the norm (e.g. 4800 samples in resting state datasets from Human Connectome Project datasets, Van Essen et al., 2013). Lastly, Smith et al. used a fixed delays of 50 [ms] in the first layer of DCM model, representing the underlying neuronal communication. This delay represents synaptic transmission delays and axonal transmission delays between nodes of the network. The constant value of delay is a crude estimation, especially given that, pairs of brain regions positioned at different distances from each other, should have different axonal transmission delays. Also, given polysynaptic connections, effective delays between neuronal populations might be much higher than the aforementioned 50 [ms], e.g.

P300 potential appears after 300 [ms] (Polich, 2007) and some other cortical potentials have even slower latencies. This lack of attention towards modelling neuronal delays, might favour non-lagged methods (i.e. Patel's tau or LiNGAM) might be preferred in this analysis over the lagged methods. Altogether, there are reasons to believe that benchmark datasets derived by Smith et al. are, to some extent, not representative of the real fMRI datasets. For these reasons, results of validation on the benchmark datasets should be interpreted with care.

There are also other, competitive generative models in the field, e.g. the model proposed by Seth et al. (2013). In this model, the authors used a simple VAR generative model in order to simulate neuronal dynamics in the testing network of 5 regions, based on work by Baccalá and Sameshima, 2001. Subsequently, the VAR model output is convolved with five different HRF kernels generated with use of the difference- of-gamma approach as implemented in SPM8 (http://www.fil.ion.ucl.ac.uk/spm/software/spm8/). Another possibility, would be to use Local Field Potentials (LFPs) instead of neuronal dynamics simulated as a system of differential equations with delay, and convolve LFPs with HRF (Deshpande, Sathian, & Hu, 2010). However, to date, Smith's synthetic datasets remains the benchmark datasets.

Furthermore, in this work, we performed the inference on the full BOLD response, without deconvolving the BOLD time series into the neuronal time series. It has been shown in synthetic and empirical data that incorporating a physiologically based model of spatiotemporal hemodynamic response function into the preprocessing pipeline leads to an improvement in the estimated neuronal activation (Aquino, Robinson, Schira, & Breakspear, 2014). It was also shown that it is generally difficult to accurately recover true task-evoked changes in BOLD fMRI time series irrespectively of the method chosen for modelling HRF response (Lindquist et al., 2009). Hence, there is a long-lasting debate in the field of connectomics on whether or not a (blind) hemodynamic deconvolution is necessary to perform the (effective) connectivity research in fMRI (Wu et al., 2013). For instance, Structural Equation Models (Mclntosh & Gonzalez-Lima, 1994) are often applied without deconvolution to fMRI datasets (Schlösser et al., 2003, Zhuang, LaConte, Peltier, Zhang, & Hu, 2005). Our previous theoretical research in synthetic datasets generated from the DCM model suggests that deconvolution is not necessary in effective connectivity research in fMRI if the method used in the study is not lag-dependent (Bielczyk et al., 2017). This is because under the assumption that the underlying signal on the neuronal level is in the low frequency range, the hemodynamic response - as a low-pass filter - does not affect the signatures of different connectivity patterns present in the outcome BOLD response. In this work, we did not perform the deconvolution step before assessing effective connectivity with any of the tested methods, including Granger Causality. In fMRI literature, GC is applied both with (David et al., 2008, Ryali, Supekar, Chen, & Menon, 2011, Ryali et al., 2016, Hutcheson et al., 2015, Wheelock et al., 2014, Sathian, Deshpande, & Stilla, 2013, Goodyear et al., 2016) and without (Zhao et al., 2016, Regner et al., 2016, Chen et al., 2017) use of deconvolution. Recent research suggests that all connectivity methods (including functional connectivity) will improve their estimation accuracy post HRF deconvolution (Rangaprakash, Wu, Marinazzo, Hu, & Deshpande, 2018). However, in this work, we chose for implementation without deconvolution, to be consistent with (Smith et al., 2011).

In this place, we would like to mention that in recent years, a lot of progress has been made in the area of modelling local hemodynamics from fMRI datasets. In example, Havlicek et al. (2011) proposed a new approach to modelling hemodynamic response functions based on cubature Kalman filtering. Furthermore, Bush et al. (2015) proposed and validated a meta-algorithm for performing semi-blind deconvolution of the BOLD  fMRI using bootstrapping. This method allows for estimating the timing of the underlying neural events stimulating BOLD responses, together with confidence levels. Sreenivasan et al. (2015) proposed a nonparametric blind BOLD deconvolution method based on homomorphic filtering.

Lastly, in our simulations we have set the connection strength to a fixed value of a = 0.9. On this stage, the output of the classifier (Eq. 4 & Eq. 5) is a binary response, i.e. an indication for a connection, either X → Y or Y → X. This indication is based on a linear combination the binary signum maps Sr, Si (Fig. 2C) with the values of the cumulants Cr, Ci computed for the given dataset X(t), Y(t), in either weighted or unweighted form. If the connection strength varies, the strength of the coupling between fractional moments in X(t) and Y(t) varies accordingly. Therefore, also the absolute values of the associated fractional cumulants will adjust. If cumulant values scale, then the RHS sum in Eq. 4, 5 will scale accordingly, but the signum of this sum should stay the same. Therefore, in noiseless case, this classifier should give the same response regardless of the connection strength $a$.

**4.2 Limitations of the method**

It is also important to remember that there are always two independent aspects to a method for causal inference. First, the method should have assumptions grounded in a biologically plausible framework relevant for the given research problem. For instance, a method for causal inference in fMRI should respect (1) the confounding, region- and subject-specific BOLD dynamics (Handwerker, Ollinger, & D'Esposito, 2004) and (2) co-occurance of cause and effect (since the time resolution of the data is low compared with the underlying neuronal dynamics; the causes and their effects most likely happen within the same frame in the fMRI data). The new methods for pairwise inference such as classifying on the basis of fractional cumulants address this issue by (1) breaking the time order, and performing causal inference on the basis of statistical properties of the distribution of the BOLD samples, and not from the timing of events; and (2) using correlation in order to detect connections. A good counterexample here is GC which has been proven useful in multiple disciplines. However, there is an ongoing discussion on whether or not GC is suited for causal interpretations of fMRI data. On the one hand, theoretical work by (Seth, Chorley, & Barnett, 2013) and (Roebroeck, Formisano, & Goebel, 2005) suggest that despite the slow hemodynamics, GC can still be informative about the directionality of causal links in the brain. Second, an estimation procedure needs to be computationally stable. Even if the generative model faithfully describes the data, it still depends on the estimation algorithm whether the method will return correct results. However, the face validity of the algorithms can only be tested in particular paradigms, in which the ground truth is known. If in the given paradigm, the ground truth is unknown, which is most often the case in fMRI experiments, only reliability can be tested.

Our approach requires certain assumptions. For instance, we assume that on the neuronal level, effects of directed connectivity are linear. This is also an assumption underlying the original DCM model used in this study. However, it is known that this is not always the case in the neuronal dynamics. For instance, shunting inhibition (Alger & Nicoll, 1979) is a phenomenon in which excitatory potential is reduced by division rather than by subtraction. However, effects such as shunting inhibition typically happen in microscale and should not affect large-scale neuronal dynamics as measured by BOLD fMRI. Therefore, we do not consider effects such as shunting inhibition as plausible confounds to our approach.

One crucial limitation of our approach as well as previous methods such as pair-wise likelihood ratios is that these techniques only retrieve the *net* connectivity. Namely, what these methods effectively pick up on is the *difference* between connectivity strengths, and in a scenario where the *connectivity* strengths X —> Y and Y —> X are equal, the outcome cumulant map for the system will have lower amplitudes than in case of a unidirectional connection X —> Y with the same connection strength. The significance of the cumulant values (whether or not the values are significantly different from zero), can be established with use of permutation testing. For more details, see: Supplementary material 6, Fig. 12B. However, since in the first step of the inference, we select only strong functional links for the further classification, we can interpret ambiguous output from the classifier as a *bidirectional* connection. Since the quality of the classifier depends on the ability to determine the directionality of unidirectional connection, we used only unidirectional connections in the validation stage. One point to make here, is that, bidirectional connections are hard to disambiguate for many methods for effective connectivity, as they represent cyclic nets. This is an issue, e.g. in applications of Bayesian nets (Pearl, 2000) in which, joint probability for a certain graph is not tractable when propagation of information in the network is cyclic. Also, the third- and fourth-cumulant method by Hyvarinen & Smith (2013) falls into this category, as the asymmetry between third cumulants in case of bidirectional connections will be equal to zero.

One interesting direction for the method development would be also classification between excitatory and inhibitory connectivity. This aspect is missing in our study, as we focus on the connections of a positive sign only (deriving a classifier able to distinguish between excitatory and inhibitory connections would require deriving an additional set of benchmark $S_r$ and $S_i$ maps built on the basis of repetitive simulations of an inhibitory connection, and creating a new set of benchmark synthetic datasets - as the original datasets by Smith et al., 2011 involve excitatory links only). The reason why inhibition is not implemented yet, is because the cumulant patterns for inhibitory connection are different from patterns given in Fig. 2 (we did not include the pictures in this manuscript though). The classifier in its current form, gives a binary decision on whether the connection is going in the direction of $X \rightarrow Y$ or $Y \rightarrow X$. The decision is based on whether the cumulant pattern obtained from the given dataset X(t), Y(t) is more similar to the signum maps derived from DCM for connection $X \rightarrow Y$ (Fig. 2 C), or to the inverse of these signum maps. In order to add inhibition to the picture, one would need to extend the number of possible classes by adding signum maps derived from DCM for inhibitory connection $X \rightarrow Y$, and introducing some metrics of distance to excitatory/inhibitory signum maps. This is the next step to take. One note on a side, is that whether or not inhibitory effective connectivity is expected in large scale networks investigated with fMRI, is a matter for debate. On one hand, it is known that long-distance projections in the brain are mostly excitatory as inhibition is typically local, presented by groups of tightly connected interneurons within single brain regions (Markram et al., 2004) - which is also modeled in the DCM generative model with use of the self-inhibition term (Friston et al., 2003). On the other hand, several anatomical and physiological studies indicate that there are also long range inhibitory connections, e.g. between hippocampus and entorhinal cortex in rats (Melzer et al., 2012).

Furthermore, we consider local variability in brain physiology by varying hemodynamic responses between realisations of the simulation, and studying the effects of the scale-free background noise on the resulting effective connectivity measures. The DCM generative model summarises the current state of knowledge about the mechanisms underlying generation of the BOLD response (Friston et al., 2017, Havlicek et al., 2015, Havlicek, Ivanov, Roebroeck, & Uludağ, 2017). Therefore, we do not have efficient ways of incorporating human brain physiology into our consideration in any more depth than this model allows for.

However, at the same time, we do not consider the influence of artefacts from the data acquisition, such as the effects of movement in the scanner. We believe that the influence of such confounders should be limited by a proper data preprocessing. For instance, motion artefacts can be reduced with use of new, data-driven protocols for motion artefacts removal such as ICA-AROMA (Pruim et al., 2015) or censoring-based artefact removal strategy based on volume censoring (Power et al., 2014). Therefore, developing efficient strategies for controlling such confounders is beyond the scope of this paper.

**4.3 Future research**

In the context of fMRI research, increasing the granularity of moments in order to better characterize the distribution is an especially useful application because the experimental fMRI datasets are short (a few thousands samples at most), therefore the estimation error for the high order integer moments of the distribution becomes high. However, the subsequent cumulants contain information redundant to a certain extent, as they are correlated for any given time series x(t). We chose the granularity which gives smooth patterns of discriminability (Fig. 3), which is $\Delta_k = 0.1$. Choosing the optimal moment resolution is a subject to future research, although we believe that increasing index resolution to substantially less than 0.1 would not be beneficial, yet it would substantially increase the computational cost for the method.

In this work, we validated our approach using synthetic benchmark datasets derived from the Dynamic Causal Modeling generative model. Using generative models is valuable in neuroimaging, in terms of validating new methods as mentioned in point 2, but also in applied research. Especially in the fields of applied computational psychiatry (Frässle et al., 2018) and network neuroscience in general (Betzel & Bassett, 2017), using generative models is valuable because these models enable inference on model parameters which afford for some degree of mechanistic interpretability on the putative processes underlying the studied phenomena. Generative models, especially Dynamic Causal Modeling, are acknowledged in multiple contexts in the field of cognitive neuroimaging, from method validation to application in clinical datasets. However, a valuable method should also give predictions testable the clinics (e.g. lead to more reliable estimation of directed causal influences during cognitive tasks, lead to better stratification of clinical populations in resting state etc.), which we will also further investigate.

In this work, we faithfully reproduced the pipeline after (Smith et al., 2011). However, since the original version of Dynamic Causal Modeling (Friston, Harrison, & Penny, 2003) based on the original Balloon-Windkessel model (Buxton, Wong, & Frank, 1998) was published, substantial advancements to the hemodynamic model have been proposed. Firstly, Obata et al., 2004 reported an error in the expression for the outcome BOLD response (Eq. 8, Supplementary Material 1: DCM forward model). Stephan, Weiskopf, Drysdale, Robinson, & Friston, 2007 proposed a more accurate expression for one of the terms in this formula. Secondly, Heinzle, Koopmans, den Ouden, Raman, & Stephan, 2016 proposed updated formula for field strengths higher than 1.5 [T]. In the following work, we will take these improvements into account.

Furthermore, the approach needs to further be tested and compared against other methods with use of experimental fMRI datasets, such as, e.g., Human Connectome Project data (van Essen et al., 2013, Barch et al., 2013). Since little is known about causal connections in large scale brain networks, especially in the resting state, such a validation might be quantitative (i.e., by means of reliability) rather than qualitative. Alternatively, one could perform such a validation in particular pathways in which one can make assumptions about the directionality of information flow, such as the dorsal and the ventral stream in the visual cortex during. One possibility is testing using datasets coming from an interventional study in which neural activity was evoked and therefore, the ground truth is known: a fused fMRI-optogenetic experiment (Ryali et al., 2016), in which intervention with use of optogenetics guarantees causal link between investigated neuronal populations.

**4.4 Conclusions**

The field of effective connectivity research in fMRI is still growing. For instance, in the recent few years, multiple algorithms for the graph network search have been developed and applied to fMRI datasets, including Independent Multiple-sample Greedy Equivalence Search (IMaGES, Ramsey et al., 2010), Group Iterative Multiple Model Estimation (GIMME, Gates & Molenaar, 2012) or Fast Greedy Equivalence Search (FGES, Ramsey et al., 2016). It is a material for a debate whether or not bivariate or multivariate inference serves a better purpose for effective connectivity research in fMRI. In our work, we focused on the pairwise inference, and we achieved a significant improvement over previous approaches for pair-wise estimation of functional causal interactions. Most importantly the robustness against known sources of variability (possible differences between up- and downstream noise magnitude and possible presence of non-Gaussian scale-free noise) significantly increases due to the simultaneous incorporation of multiple aspects of the associated BOLD distributions. We believe that, as this approach based on fractional moments of a distribution increases resilience of the methods for pairwise connectivity to potential confounds in the experimental data, it can become a generic method to increase the power of causal discovery studies, both in cognitive neuroimaging and beyond.

**References**

Albantakis, L., Marshall, W., Hoel, E., & Tononi, G. (2017). What caused what? A quantitative account of actual causation using dynamical causal networks. *ArXiv:1708.06716 [Cs, Math, Stat]*. Retrieved from http://arxiv.org/abs/1708.06716

Aldrich, J. (1995). Correlations Genuine and Spurious in Pearson and Yule. *Statistical Science*, *10*(4), 364–376. https://doi.org/10.1214/ss/1177009870

Alger, B. E., & Nicoll, R. A. (1979). GABA-mediated biphasic inhibitory responses in hippocampus | Nature. *Nature*, *281*, 315–317. https://doi.org/10.1038/281315a0

Altman, N., & Krzywinski, M. (2015). Points of Significance: Association, correlation and causation. *Nature Methods*, *12*, 899–900. https://doi.org/10.1038/nmeth.3587

Aquino, K. M., Robinson, P. A., Schira, M. M., & Breakspear, M. (2014). Deconvolution of neural dynamics from fMRI data using a spatiotemporal hemodynamic response function. *NeuroImage*, *94*, 203–215. https://doi.org/10.1016/j.neuroimage.2014.03.001

Arichi, T., Fagiolo, G., Varela, M., Melendez-Calderon, A., Allievi, A., Merchant, N., … Edwards, A. D. (2012). Development of BOLD signal hemodynamic responses in the human brain. *NeuroImage*, *63*(2), 663–673. https://doi.org/10.1016/j.neuroimage.2012.06.054

Arioli, M., Perani, D., Cappa, S., Proverbio, A. M., Zani, A., Falini, A., & Canessa, N. (2018). Affective and cooperative social interactions modulate effective connectivity within and between the mirror and mentalizing systems. *Human Brain Mapping*, *39*(3), 1412–1427. https://doi.org/10.1002/hbm.23930

Baccalá, L. A., & Sameshima, K. (2001). Partial directed coherence: a new concept in neural structure determination. *Biological Cybernetics*, *84*(6), 463–474. https://doi.org/10.1007/PL00007990

Barch, D. M., Burgess, G. C., Harms, M. P., Petersen, S. E., Schlaggar, B. L., Corbetta, M., … WU-Minn HCP Consortium. (2013). Function in the human connectome: task-fMRI and individual differences in behavior. *NeuroImage*, *80*, 169–189. https://doi.org/10.1016/j.neuroimage.2013.05.033

Barnett, L., & Seth, A. K. (2014). The MVGC multivariate Granger causality toolbox: a new approach to Granger-causal inference. *Journal of Neuroscience Methods*, *223*, 50–68. https://doi.org/10.1016/j.jneumeth.2013.10.018

Barnett, L., & Seth, A. K. (2015). Granger causality for state-space models. *Physical Review E*, *91*(4), 040101. https://doi.org/10.1103/PhysRevE.91.040101

Bédard, C., Kröger, H., & Destexhe, A. (2006). Does the 1/f frequency scaling of brain signals reflect self-organized critical states? *Physical Review Letters*, *97*(11), 118102. https://doi.org/10.1103/PhysRevLett.97.118102

Berkson, J. (1946). Limitations of the application of fourfold table analysis to hospital data. *Biometrics Bulletin*, *2*(3), 47–53. https://doi.org/10.2307/3002000

Bielczyk, N. Z., Buitelaar, J. K., Glennon, J. C., & Tiesinga, P. H. E. (2015). Circuit to Construct Mapping: A Mathematical Tool for Assisting the Diagnosis and Treatment in Major Depressive Disorder. *Frontiers in Psychiatry*, *6*. https://doi.org/10.3389/fpsyt.2015.00029

Bielczyk, N. Z., Llera, A., Buitelaar, J. K., Glennon, J. C., & Beckmann, C. F. (2017). The impact of hemodynamic variability and signal mixing on the identifiability of effective connectivity structures in BOLD fMRI. *Brain and Behavior*, *7*(8). https://doi.org/10.1002/brb3.777

Bielczyk, N. Z., Walocha, F., Ebel, P. W., Haak, K. V., Llera, A., Buitelaar, J. K., … Beckmann, C. F. (2018). Thresholding functional connectomes by means of mixture modeling. *NeuroImage*, *171*, 402–414. https://doi.org/10.1016/j.neuroimage.2018.01.003

Bielczyk, N. Z., Uithol, S., van Mourik, T., Anderson, P., Glennon, J. C., & Buitelaar, J. K. (2018). Disentangling casual webs in the brain using functional magnetic resonance imaging: A review of current approaches. *Network Neuroscience*, 1–37. https://doi.org/10.1162/netn_a_00062

Bush, K., Cisler, J., Bian, J., Hazaroglu, G., Hazaroglu, O., & Kilts, C. (2015). Improving the Precision of fMRI BOLD Signal Deconvolution with Implications for Connectivity Analysis. *Magnetic Resonance Imaging*, *33*(10), 1314–1323. https://doi.org/10.1016/j.mri.2015.07.007

Buxton, R. B., Wong, E. C., & Frank, L. R. (1998). Dynamics of blood flow and oxygenation changes during brain activation: the balloon model. *Magnetic Resonance in Medicine*, *39*(6), 855–864. https://doi.org/10.1002/mrm. 1910390602

Chen, Y.-C., Xia, W., Chen, H., Feng, Y., Xu, J.-J., Gu, J.-P., … Yin, X. (2017). Tinnitus distress is linked to enhanced resting-state functional connectivity from the limbic system to the auditory cortex. *Human Brain Mapping*, *38*(5), 2384–2397. https://doi.org/10.1002/hbm.23525

Chicharro, D. (2011). On the spectral formulation of Granger causality. *Biological Cybernetics*, *105*(5–6), 331–347. https://doi.org/10.1007/s00422-011-0469-z

David, O., & Friston, K. J. (2003). A neural mass model for MEG/EEG: *NeuroImage*, *20*(3), 1743–1755. https://doi.org/10.1016/j.neuroimage.2003.07.015

David, O., Guillemain, I., Saillet, S., Reyt, S., Deransart, C., Segebarth, C., & Depaulis, A. (2008). Identifying Neural Drivers with Functional MRI: An Electrophysiological Validation. *PLoS Biology*, *6*(12). https://doi.org/10.1371/journal.pbio.0060315

Deco, G., Jirsa, V. K., & McIntosh, A. R. (2011). Emerging concepts for the dynamical organization of resting-state activity in the brain. *Nature Reviews. Neuroscience*, *12*(1), 43–56. https://doi.org/10.1038/nrn2961

Dehghani, N., Bédard, C., Cash, S. S., Halgren, E., & Destexhe, A. (2010). Comparative power spectral analysis of simultaneous elecroencephalographic and magnetoencephalographic recordings in humans suggests non-resistive extracellular media. *Journal of Computational Neuroscience*, *29*(3), 405–421. https://doi.org/10.1007/s10827-010-0263-2

Deshpande, G., Sathian, K., & Hu, X. (2010). Effect of Hemodynamic Variability on Granger Causality Analysis of fMRI. *NeuroImage*, *52*(3), 884–896. https://doi.org/10.1016/j.neuroimage.2009.11.060

Di, X., & Biswal, B. B. (2019). Toward Task Connectomics: Examining Whole-Brain Task Modulated Connectivity in Different Task Domains. *Cerebral Cortex*, *29*(4), 1572–1583. https://doi.org/10.1093/cercor/bhy055

Devonshire, I. M., Papadakis, N. G., Port, M., Berwick, J., Kennerley, A. J., Mayhew, J. E. W., & Overton, P. G. (2012). Neurovascular coupling is brain region-dependent. *NeuroImage*, *59*(3), 1997–2006. https://doi.org/10.1016/j.neuroimage.2011.09.050

Dremlin, I. M. (1994). Fractional moments of distributions. *Journal of Experimental and Theoretical Physics Letters*, *59*(9), 585–588.

Van Essen, D. C., Smith, S. M., Barch, D. M., Behrens, T. E. J., Yacoub, E., Ugurbil, K., & WU-Minn HCP Consortium. (2013). The WU-Minn Human Connectome Project: an overview. *NeuroImage*, *80*, 62–79. https://doi.org/10.1016/j.neuroimage.2013.05.041

Faes, L., Erla, S., Porta, A., & Nollo, G. (2013). A framework for assessing frequency domain causality in physiological time series with instantaneous effects. *Philosophical Transactions. Series A,*

*Mathematical, Physical, and Engineering Sciences*, *371*(1997), 20110618. https://doi.org/10.1098/rsta.2011.0618

Frässle, S., Yao, Y., Schöbi, D., Aponte, E. A., Heinzle, J., & Stephan, K. E. (2018). Generative models for clinical applications in computational psychiatry. *Wiley Interdisciplinary Reviews. Cognitive Science*, *9*(3), e1460. https://doi.org/10.1002/wcs.1460

Friedman, J., Hastie, T., & Tibshirani, R. (2008). Sparse inverse covariance estimation with the graphical lasso. *Biostatistics (Oxford, England)*, *9*(3), 432–441. https://doi.org/10.1093/biostatistics/kxm045

Friston, K. J., Harrison, L., & Penny, W. (2003). Dynamic causal modelling. *NeuroImage*, *19*(4), 1273–1302.

Friston, K. J. (2011). Functional and effective connectivity: a review. *Brain Connectivity*, *1*(1), 13–36. https://doi.org/10.1089/brain.2011.0008

Friston, K. J., Preller, K. H., Mathys, C., Cagnan, H., Heinzle, J., Razi, A., & Zeidman, P. (2017). Dynamic causal modelling revisited. *NeuroImage*. https://doi.org/10.1016/j.neuroimage.2017.02.045

Gates, K. M., & Molenaar, P. C. M. (2012). Group search algorithm recovers effective connectivity maps for individuals in homogeneous and heterogeneous samples. *NeuroImage*, *63*(1), 310–319. https://doi.org/10.1016/j.neuroimage.2012.06.026

Geweke, J. (1982). Measurement of Linear Dependence and Feedback between Multiple Time Series. *Journal of the American Statistical Association*, *77*(378), 304–313. https://doi.org/10.1080/01621459.1982.10477803

Geweke, J. F. (1984). Measures of Conditional Linear Dependence and Feedback between Time Series. *Journal of the American Statistical Association*, *79*(388), 907–915. https://doi.org/10.1080/01621459.1984.10477110

Goodyear, K., Parasuraman, R., Chernyak, S., Madhavan, P., Deshpande, G., & Krueger, F. (2016). Advice Taking from Humans and Machines: An fMRI and Effective Connectivity Study. *Frontiers in Human Neuroscience*, *10*, 542. https://doi.org/10.3389/fnhum.2016.00542

Granger, C. W. J. (1969). Investigating Causal Relations by Econometric Models and Cross-spectral Methods. *Econometrica*, *37*(3), 424–438. https://doi.org/10.2307/1912791

Hamilton, F., Berry, T., Peixoto, N., & Sauer, T. (2013). Real-time tracking of neuronal network structure using data assimilation. *Physical Review E*, *88*(5). https://doi.org/10.1103/PhysRevE.88.052715

Handwerker, D. A., Ollinger, J. M., & D'Esposito, M. (2004). Variation of BOLD hemodynamic responses across subjects and brain regions and their effects on statistical analyses. *NeuroImage*, *21*(4), 1639–1651. https://doi.org/10.1016/j.neuroimage.2003.11.029

Havlicek, M., Friston, K. J., Jan, J., Brazdil, M., & Calhoun, V. D. (2011). Dynamic modeling of neuronal responses in fMRI using cubature Kalman filtering. *NeuroImage*, *56*(4), 2109–2128. https://doi.org/10.1016/j.neuroimage.2011.03.005

Havlicek, M., Roebroeck, A., Friston, K., Gardumi, A., Ivanov, D., & Uludag, K. (2015). Physiologically informed dynamic causal modeling of fMRI data. *NeuroImage*, *122*, 355–372. https://doi.org/10.1016/j.neuroimage.2015.07.078

Havlicek, M., Ivanov, D., Roebroeck, A., & Uludağ, K. (2017). Determining Excitatory and Inhibitory Neuronal Activity from Multimodal fMRI Data Using a Generative Hemodynamic Model. *Frontiers in Neuroscience*, *11*, 616. https://doi.org/10.3389/fnins.2017.00616

He, B. J. (2014). Scale-free brain activity: past, present, and future. *Trends in Cognitive Sciences*, *18*(9), 480–487. https://doi.org/10.1016/j.tics.2014.04.003

Heinzle, J., Koopmans, P. J., den Ouden, H. E. M., Raman, S., & Stephan, K. E. (2016). A hemodynamic model for layered BOLD signals. *NeuroImage*, *125*, 556–570. https://doi.org/10.1016/j.neuroimage.2015.10.025

van den Heuvel, M. P., de Lange, S. C., Zalesky, A., Seguin, C., Yeo, B. T. T., & Schmidt, R. (2017). Proportional thresholding in resting-state fMRI functional connectivity networks and consequences for patient-control connectome studies: Issues and recommendations. *NeuroImage*, *152*, 437–449. https://doi.org/10.1016/j.neuroimage.2017.02.005

Hindmarsh, J. L., & Rose, R. M. (1984). A model of neuronal bursting using three coupled first order differential equations. *Proceedings of the Royal Society of London. Series B, Biological Sciences*, *221*(1222), 87–102. https://doi.org/10.1098/rspb.1984.0024

Hinne, M., Janssen, R. J., Heskes, T., & van Gerven, M. A. J. (2015). Bayesian Estimation of Conditional Independence Graphs Improves Functional Connectivity Estimates. *PLoS Computational Biology*, *11*(11), e1004534. https://doi.org/10.1371/journal.pcbi.1004534

Hodgkin, A. L., & Huxley, A. F. (1952). A quantitative description of membrane current and its application to conduction and excitation in nerve. *The Journal of Physiology*, *117*(4), 500–544.

Hume, D. (1772). Cause and Effect. In *An Enquiry Concerning Human Understanding*.

Hutcheson, N. L., Sreenivasan, K. R., Deshpande, G., Reid, M. A., Hadley, J., White, D. M., … Lahti, A. C. (2015). Effective connectivity during episodic memory retrieval in schizophrenia participants before and after antipsychotic medication. *Human Brain Mapping*, *36*(4), 1442–1457. https://doi.org/10.1002/hbm.22714

Hyvärinen, A., Zhang, K., Shimizu, S., & Hoyer, P. O. (2010). Estimation of a Structural Vector Autoregression Model Using Non-Gaussianity. *Journal of Machine Learning Research*, *11*(May), 1709–1731.

Hyvärinen, A., & Smith, S. M. (2013). Pairwise Likelihood Ratios for Estimation of Non-Gaussian Structural Equation Models. *Journal of Machine Learning Research*, *14*(Jan), 111–152.

Janzing, D., Mooij, J., Zhang, K., Lemeire, J., Zscheischler, J., Daniušis, P., … Schölkopf, B. (2012). Information-geometric approach to inferring causal directions. *Artificial Intelligence*, *182–183*, 1–31. https://doi.org/10.1016/j.artint.2012.01.002

Ji, J., Liu, J., Liang, P., & Zhang, A. (2016). Learning Effective Connectivity Network Structure from fMRI Data Based on Artificial Immune Algorithm. *PloS One*, *11*(4), e0152600. https://doi.org/10.1371/journal.pone.0152600

Kamiński, M. J., & Blinowska, K. J. (1991). A new method of the description of the information flow in the brain structures. *Biological Cybernetics*, *65*(3), 203–210.

Kennaway, R. (2015). When causation does not imply correlation: robust violations of the Faithfulness axiom. *ArXiv:1505.03118 [Math, Stat]*. Retrieved from http://arxiv.org/abs/1505.03118

Kiebel, S. J., Garrido, M. I., Moran, R., Chen, C.-C., & Friston, K. J. (2009). Dynamic causal modeling for EEG and MEG. *Human Brain Mapping*, *30*(6), 1866–1876. https://doi.org/10.1002/hbm.20775

Kötter, R., & Stephan, K. E. (2003). Network participation indices: characterizing component roles for information processing in neural networks. *Neural Networks*, *16*(9), 1261–1275. https://doi.org/10.1016/j.neunet.2003.06.002

Kruger, P. (2018). Why it is not a ‘failure’ to leave academia. *Nature*, *560*, 7716. https://doi.org/10.1038/d41586-018-05838-y

Ledoit, O., & Wolf, M. (2004). A Well-conditioned Estimator for Large-dimensional Covariance Matrices. *J. Multivar. Anal.*, *88*(2), 365–411. https://doi.org/10.1016/S0047-259X(03)00096-4

Lin, T. W., Das, A., Krishnan, G. P., Bazhenov, M., & Sejnowski, T. J. (2017). Differential Covariance: A New Class of Methods to Estimate Sparse Connectivity from Neural Recordings. *Neural Computation*, *29*(10), 2581–2632. https://doi.org/10.1162/neco_a_01008

Lindquist, M. A., Loh, J. M., Atlas, L. Y., & Wager, T. D. (2009). Modeling the Hemodynamic Response Function in fMRI: Efficiency, Bias and Mis-modeling. *Neuroimage*, *45*(1 Suppl), S187–S198. https://doi.org/10.1016/j.neuroimage.2008.10.065

Markram, H., Toledo-Rodriguez, M., Wang, Y., Gupta, A., Silberberg, G., & Wu, C. (2004). Interneurons of the neocortical inhibitory system. *Nature Reviews. Neuroscience*, *5*(10), 793–807. https://doi.org/10.1038/nrn1519

Marrelec, G., Krainik, A., Duffau, H., Pélégrini-Issac, M., Lehéricy, S., Doyon, J., & Benali, H. (2006). Partial correlation for functional brain interactivity investigation in functional MRI. *NeuroImage*, *32*(1), 228–237. https://doi.org/10.1016/j.neuroimage.2005.12.057

Matsui, M., & Pawlas, Z. (2013). Fractional absolute moments of heavy tailed distributions. *ArXiv:1301.4804 [Math, Stat]*. Retrieved from http://arxiv.org/abs/1301.4804

McLntosh, A. R., & Gonzalez−Lima, F. (1994). Structural equation modeling and its application to network analysis in functional brain imaging. *Human Brain Mapping*, *2*(1–2), 2–22. https://doi.org/10.1002/hbm.460020104

Melzer, S., Michael, M., Caputi, A., Eliava, M., Fuchs, E. C., Whittington, M. A., & Monyer, H. (2012). Long-Range–Projecting GABAergic Neurons Modulate Inhibition in Hippocampus and Entorhinal Cortex. *Science*, *335*(6075), 1506–1510. https://doi.org/10.1126/science.1217139

Moran, R. J., Pinotsis, D. A., & Friston, K. J. (2013). Neural masses and fields in dynamic causal modeling. *Frontiers in Computational Neuroscience*, *7*. https://doi.org/10.3389/fncom.2013.00057

Nackaerts, E., Michely, J., Heremans, E., Swinnen, S. P., Smits-Engelsman, B. C. M., Vandenberghe, W., … Nieuwboer, A. (2018). Training for Micrographia Alters Neural Connectivity in Parkinson's Disease. *Frontiers in Neuroscience*, *12*, 3. https://doi.org/10.3389/fnins.2018.00003

Nie, L., Yang, X., Matthews, P. M., Xu, Z., & Guo, Y. (2015). Minimum Partial Correlation: An Accurate and Parameter-Free Measure of Functional Connectivity in fMRI. In Y. Guo, K. Friston, F. Aldo, S. Hill, & H. Peng (Eds.), *Brain Informatics and Health* (Vol. 9250, pp. 125–134). Cham: Springer International Publishing.

Obata, T., Liu, T. T., Miller, K. L., Luh, W.-M., Wong, E. C., Frank, L. R., & Buxton, R. B. (2004). Discrepancies between BOLD and flow dynamics in primary and supplementary motor areas: application of the balloon model to the interpretation of BOLD transients. *NeuroImage*, *21*(1), 144–153. https://doi.org/10.1016/j.neuroimage.2003.08.040

Patel, R. S., Bowman, F. D., & Rilling, J. K. (2006). A Bayesian approach to determining connectivity of the human brain. *Human Brain Mapping*, *27*(3), 267–276. https://doi.org/10.1002/hbm.20182

Pearl, J. (2000). *Causality: Models, Reasoning, and Inference*. Cambridge University Press.

Pinotsis, D. A., Geerts, J. P., Pinto, L., FitzGerald, T. H. B., Litvak, V., Auksztulewicz, R., & Friston, K. J. (2017). Linking canonical microcircuits and neuronal activity: Dynamic causal modelling of laminar recordings. *NeuroImage*, *146*, 355–366. https://doi.org/10.1016/j.neuroimage.2016.11.041

Polich, J. (2007). Updating P300: An Integrative Theory of P3a and P3b. *Clinical Neurophysiology : Official Journal of the International Federation of Clinical Neurophysiology*, *118*(10), 2128–2148. https://doi.org/10.1016/j.clinph.2007.04.019

Pool, E.-M., Leimbach, M., Binder, E., Nettekoven, C., Eickhoff, S. B., Fink, G. R., & Grefkes, C. (2018). Network dynamics engaged in the modulation of motor behavior in stroke patients. *Human Brain Mapping*, *39*(3), 1078–1092. https://doi.org/10.1002/hbm.23872

Power, J. D., Mitra, A., Laumann, T. O., Snyder, A. Z., Schlaggar, B. L., & Petersen, S. E. (2014). Methods to detect, characterize, and remove motion artifact in resting state fMRI. *NeuroImage*, *84*. https://doi.org/10.1016/j.neuroimage.2013.08.048

Pruim, R. H. R., Mennes, M., van Rooij, D., Llera, A., Buitelaar, J. K., & Beckmann, C. F. (2015). ICA-AROMA: A robust ICA-based strategy for removing motion artifacts from fMRI data. *Neuro-Image*, *112*, 267–277. https://doi.org/10.1016/j.neuroimage.2015.02.064

Ramsey, J. D., Hanson, S. J., Hanson, C., Halchenko, Y. O., Poldrack, R. A., & Glymour, C. (2010). Six problems for causal inference from fMRI. *NeuroImage*, *49*(2), 1545–1558. https://doi.org/10.1016/j.neuroimage.2009.08.065

Rangaprakash, D., Wu, G.-R., Marinazzo, D., Hu, X., & Deshpande, G. (2018). Hemodynamic response function (HRF) variability confounds resting-state fMRI functional connectivity. *Magnetic Resonance in Medicine*, *80*(4), 1697–1713. https://doi.org/10.1002/mrm.27146

Regner, M. F., Saenz, N., Maharajh, K., Yamamoto, D. J., Mohl, B., Wylie, K., … Tanabe, J. (2016). Top-Down Network Effective Connectivity in Abstinent Substance Dependent Individuals. *PLOS ONE*, *11*(10), e0164818. https://doi.org/10.1371/journal.pone.0164818

Roebroeck, A., Formisano, E., & Goebel, R. (2005). Mapping directed influence over the brain using Granger causality and fMRI. *NeuroImage*, *25*(1), 230–242. https://doi.org/10.1016/j.neuroimage.2004.11.017

Ryali, S., Supekar, K., Chen, T., & Menon, V. (2011). Multivariate dynamical systems models for estimating causal interactions in fMRI. *NeuroImage*, *54*(2), 807–823. https://doi.org/10.1016/j.neuroimage.2010.09.052

Ryali, S., Shih, Y.-Y. I., Chen, T., Kochalka, J., Albaugh, D., Fang, Z., … Menon, V. (2016). Combining optogenetic stimulation and fMRI to validate a multivariate dynamical systems model for estimating causal brain interactions. *NeuroImage*, *132*, 398–405. https://doi.org/10.1016/j.neuroimage.2016.02.067

Sathian, K., Deshpande, G., & Stilla, R. (2013). Neural Changes with Tactile Learning Reflect Decision-Level Reweighting of Perceptual Readout. *Journal of Neuroscience*, *33*(12), 5387–5398. https://doi.org/10.1523/JNEUROSCI.3482-12.2013

Schlösser, R., Gesierich, T., Kaufmann, B., Vucurevic, G., Hunsche, S., Gawehn, J., & Stoeter, P. (2003). Altered effective connectivity during working memory performance in schizophrenia: a study with fMRI and structural equation modeling. *NeuroImage*, *19*(3), 751–763.

Schurger, A., & Uithol, S. (2015). Nowhere and Everywhere: The Causal Origin of Voluntary Action. *Review of Philosophy and Psychology*, *6*(4), 761–778.

Seth, A. K., Chorley, P., & Barnett, L. C. (2013). Granger causality analysis of fMRI BOLD signals is invariant to hemodynamic convolution but not downsampling. *NeuroImage*, *65*, 540–555. https://doi.org/10.1016/j.neuroimage.2012.09.049

Seth, A. K., Barrett, A. B., & Barnett, L. (2015). Granger causality analysis in neuroscience and neuroimaging. *The Journal of Neuroscience: The Official Journal of the Society for Neuroscience*, *35*(8), 3293–3297. https://doi.org/10.1523/JNEUROSCI.4399-14.2015

Sgouritsa, E., Janzing, D., Hennig, P., & Schölkopf, B. (2015). Inference of Cause and Effect with Unsupervised Inverse Regression. In *Proceedings of the UAI 2015 Conference on Advances in Causal Inference - Volume 1504* (pp. 89–89). Aachen, Germany, Germany: CEUR-WS.org. Retrieved from http://dl.acm.org/citation.cfm?id=3020267.3020280

Shimizu, S., Hoyer, P. O., Hyvärinen, A., & Kerminen, A. (2006). A Linear Non-Gaussian Acyclic Model for Causal Discovery. *J. Mach. Learn. Res.*, *7*, 2003–2030.

Silver, R. A. (2010). Neuronal arithmetic. *Nature Reviews Neuroscience*, *11*(7), 474–489. https://doi.org/10.1038/nrn2864

Smith, S. M., Miller, K. L., Salimi-Khorshidi, G., Webster, M., Beckmann, C. F., Nichols, T. E., … Woolrich, M. W. (2011). Network modelling methods for FMRI. *NeuroImage*, *54*(2), 875–891. https://doi.org/10.1016/j.neuroimage.2010.08.063

Steen, F. V. de, Almgren, H. B. J., Razi, A., Friston, K. J., & Marinazzo, D. (2018). Dynamic causal modelling of fluctuating connectivity in resting-state EEG. *BioRxiv*, 303933. https://doi.org/10.1101/303933

Stephan, K. E., Weiskopf, N., Drysdale, P. M., Robinson, P. A., & Friston, K. J. (2007). Comparing hemodynamic models with DCM. *NeuroImage*, *38*(3), 387–401. https://doi.org/10.1016/j.neuroimage.2007.07.040

Stephan, K. E., Kasper, L., Harrison, L. M., Daunizeau, J., den Ouden, H. E. M., Breakspear, M., & Friston, K. J. (2008). Nonlinear Dynamic Causal Models for fMRI. *NeuroImage*, *42*(2), 649–662. https://doi.org/10.1016/j.neuroimage.2008.04.262

Stokes, P. A., & Purdon, P. L. (2017). A study of problems encountered in Granger causality analysis from a neuroscience perspective. *Proceedings of the National Academy of Sciences of the United States of America*, *114*(34), E7063–E7072. https://doi.org/10.1073/pnas.1704663114

Valdes-Sosa, P. A., Roebroeck, A., Daunizeau, J., & Friston, K. (2011). Effective connectivity: influence, causality and biophysical modeling. *NeuroImage*, *58*(2), 339–361. https://doi.org/10.1016/j.neuroimage.2011.03.058

Varoquaux, G., & Craddock, R. C. (2013). Learning and comparing functional connectomes across subjects. *NeuroImage*, *80*, 405–415. https://doi.org/10.1016/j.neuroimage.2013.04.007

Wang, H. E., Bénar, C. G., Quilichini, P. P., Friston, K. J., Jirsa, V. K., & Bernard, C. (2014). A systematic framework for functional connectivity measures. *Frontiers in Neuroscience*, *8*. https://doi.org/10.3389/fnins.2014.00405

Wang, Y., David, O., Hu, X., & Deshpande, G. (2017). Can Patel's τ accurately estimate directionality of connections in brain networks from fMRI? *Magnetic Resonance in Medicine*, *78*(5), 2003–2010. https://doi.org/10.1002/mrm.26583

Wheelock, M. D., Sreenivasan, K. R., Wood, K. H., Ver Hoef, L. W., Deshpande, G., & Knight, D. C. (2014). Threat-related learning relies on distinct dorsal prefrontal cortex network connectivity. *NeuroImage*, *102 Pt 2*, 904–912. https://doi.org/10.1016/j.neuroimage.2014.08.005

Wong, K.-F., & Wang, X.-W. (2006). A Recurrent Network Mechanism of Time Integration in Perceptual Decisions. *Journal of Neuroscience*, *26*(4), 1314–1328. https://doi.org/10.1523/JNEUROSCI.3733-05.2006

Wu, G.-R., Liao, W., Stramaglia, S., Ding, J.-R., Chen, H., & Marinazzo, D. (2013). A blind deconvolution approach to recover effective connectivity brain networks from resting state fMRI data. *Medical Image Analysis*, *17*(3), 365–374. https://doi.org/10.1016/j.media.2013.01.003

Zhang, L., Opmeer, E. M., van der Meer, L., Aleman, A., Ćurčić-Blake, B., & Ruhé, H. G. (2018). Altered frontal-amygdala effective connectivity during effortful emotion regulation in bipolar disorder. *Bipolar Disorders*, *20*(4), 349–358. https://doi.org/10.1111/bdi.12611

Zhao, Z., Wang, X., Fan, M., Yin, D., Sun, L., Jia, J., … Gong, J. (2016). Altered Effective Connectivity of the Primary Motor Cortex in Stroke: A Resting-State fMRI Study with Granger Causality Analysis. *PloS One*, *11*(11), e0166210. https://doi.org/10.1371/journal.pone.0166210

Zhuang, J., LaConte, S., Peltier, S., Zhang, K., & Hu, X. (2005). Connectivity exploration with structural equation modeling: an fMRI study of bimanual motor coordination. *NeuroImage*, *25*(2), 462–470. https://doi.org/10.1016/j.neuroimage.2004.11.007

**Disclosure/Conflict-of-Interest Statement**

The authors declare that the research was conducted in the absence of any commercial or financial relation-ships that could be construed as a potential conflict of interest. JCG has acted as a consultant to Boehringer Ingelheim in the last 4 years, but is not an employee or shareholder of this company.

**Author Contributions**

NZB came up with the concept of fractional moments and cumulants of the BOLD distribution (Eqs. 1 & 2), and designed the classifier to combine fractional cumulants (Eqs. 3 & 4). NZB reproduced the study by Smith et al. (2011) to create synthetic datasets for this study. NZB, AL and CFB designed the pipeline to validate the classifier in naturalistic conditions (i. e. with increased levels of background noise and uneven SNRs). NZB conducted the study and drafted the manuscript. CFB, AL and JKB critically revised the work.

**Acknowledgments**

We would like to thank to the whole SIN group at the Donders Institute for Brain, Cognition and Behavior for advice and engagement in the project.

**Funding**

JCG has received funding from the European Community's Seventh Framework Programme (FP7/2007-2013) under grant agreement no 305697 (OPTIMISTIC), the European Community's Seventh Framework Programme (FP7/2007-2013) under grant agreement no 278948 (TACTICS) and European Union's Seventh Framework Programme for research, technological development and demonstration under grant agreement no 603016 (MATRICS). CFB gratefully acknowledges funding from the Wellcome Trust UK Strategic Award [098369/Z/12/Z]. CFB is supported by the Netherlands Organisation for Scientific Research (NWO-Vidi 864-12-003).

## Supplemental Materials

**Supplementary material 1** DCM forward model

In this work, we use the original, single node per region DCM (Friston et al., 2003, Smith et al., 2011). This model operationalizes the generation of BOLD response from the neuronal networks across two levels: non-observable neuronal level and the observable hemodynamic level.

The latent neuronal dynamics is described by the simple differential relationship:

$$\frac{d\vec{z}}{dt} = A\vec{z}(t-\tau) + C\vec{u}(t) + \vec{\sigma}(t) \quad (6)$$

where z denotes the temporary activity across all nodes, u(t) denotes binary inputs (trains of on- and off- states in our case), A denotes the adjacency matrix of effective connectivity and C denotes connections from (experimental) inputs to the nodes, denotes a lag in the neuronal communication, and σ(t) denotes the level of stochasticity on the neuronal level. In our network setup, the modulatory connectivity does not play a role for the research question, therefore we set all the modulatory connections B from the original DCM model (Friston et al., 2003) to zero. The connectivity (a.k.a. adjacency) matrix A contains self-inhibition in every node as originally proposed in (Friston et al., 2003). Additionally, we use small, biologically plausible time lags of 50 [ms] in the communication between areas as also implemented in (Smith et al., 2011), therefore the simulated network becomes a system of delayed differential equations in fact.

In this context, the stochastic term σ(t) represents neuronal innovations which are not the part of the communication between nodes of the investigated network. It can either represent intrinsic dynamics in the given node (other than inhibition), or input from areas outside the investigated network. Strictly speaking, these innovations are not a 'noise' (which would mean stochasticity added to the neuronal time series on the top of the simulated dynamics), but rather a background neuronal dynamics which cannot be explained by the given model. However, for the sake of simplicity we will refer to σ(t) as noise in the text below.

The observational level is given by the classic model for the hemodynamic response, referred to as Balloon-Windkessel model (Buxton et al., 1998, Friston et al., 2003), is described node-wide, and for every node *i* it is described by the dynamics of four biophysiological variables as follows:

$$\frac{ds_i}{dt} = z_i(t) - \kappa_i s_i(t) - \gamma_i(f_i(t) - 1)$$

$$\frac{df_i}{dt} = s_i(t) \qquad (7)$$

$$\lambda_i \frac{dv_i}{dt} = f_i(t) - v_i^{1/\alpha}(t)$$

$$\lambda_i \frac{dq_i(t)}{dt} = f_i(t) \frac{E(f_i(t), \rho_i)}{\rho_i} - v_i^{\frac{1}{\alpha}-1}(t) q_i(t)$$

where $s_i(t)$ - vasodilatory signal, $f_i(t)$ - inflow, $v_i(t)$ - blood volume, $q_i(t)$ - deoxyhemoglobin content, $E(f, \rho) = 1 - (1 - \rho)^{1/f}$. The model involves five node-specific constants: $\kappa$ - rate of signal decay, $\gamma$ - rate of flow-dependent elimination, $\lambda$ - hemodynamic transit time, $\alpha$ - Grubb's exponent, $\rho$ - resting oxygen extraction fraction. Then, the following expression describes the outcome BOLD response:

$$y(t) = V_0\Big(7\rho_i\big(1 - q_i(t)\big)\Big) + 2\left(1 - \frac{q_i(t)}{v_i(t)}\right) + (2\rho_i - 0.2)(1 - v_i(t)) \quad (8)$$

Inputs to the network were simulated as in [Smith et al., 2011]: as independent trains of on- and off-states with time resolution of TR = 5 [ms]. The probability of state switches was governed by a Poissonian process of a mean on-state duration of 2.5 [s], and a mean off- state duration of 7.5 [s].

**Supplementary material 2** Benchmark synthetic datasets

Benchmark datasets by Smith et al., 2011 are build using the DCM forward model described in Supplementary Material 1. All the variations of the connectivity used in this study are acyclic and sparse, and only the size varies between 5 and 50 nodes.

The DCM forward model allows for emulating a variety of experimental conditions, such as the number of nodes in the network (N), the session duration (SD), the time resolution of the data (TR), the amount of thermal noise added to the BOLD response or the variability in delay of the hemody-namic response. In some simulations, additional features were introduced, e.g., shared inputs or backward connections. All the parameters are summarized in Table 1.

| no | N | S D (min) | TR (s) | noise (%) | std of the HRF | Other features |
|---|---|---|---|---|---|---|
| 1 | 5 | 10 | 3.0 | 1.0 | 0.5 | |
| 2 | 10 | 10 | 3.0 | 1.0 | 0.5 | |
| 3 | 15 | 10 | 3.0 | 1.0 | 0.5 | |
| 4 | 50 | 10 | 3.0 | 1.0 | 0.5 | |
| 5 | 5 | 60 | 3.0 | 1.0 | 0.5 | |
| 6 | 10 | 60 | 3.0 | 1.0 | 0.5 | |
| 7 | 5 | 250 | 3.0 | 1.0 | 0.5 | |
| 8 | 5 | 10 | 3.0 | 1.0 | 0.5 | shared inputs |
| 9 | 5 | 250 | 3.0 | 1.0 | 0.5 | shared inputs |
| 10 | 5 | 10 | 3.0 | 1.0 | 0.5 | global mean confound |
| 11 | 10 | 10 | 3.0 | 1.0 | 0.5 | bad ROIs (time series mixedwith each other) |
| 12 | 10 | 10 | 3.0 | 1.0 | 0.5 | bad ROIs (new random time series mixed in) |
| 13 | 5 | 10 | 3.0 | 1.0 | 0.5 | backwards connections |
| 14 | 5 | 10 | 3.0 | 1.0 | 0.5 | cyclic connections |
| 15 | 5 | 10 | 3.0 | 0.1 | 0.5 | stronger connections |
| 16 | 5 | 10 | 3.0 | 1.0 | 0.5 | more connections |
| 17 | 10 | 10 | 3.0 | 0.1 | 0.5 | |
| 18 | 5 | 10 | 3.0 | 1.0 | 0.0 | |
| 19 | 5 | 10 | 0.25 | 0.1 | 0.5 | neural lag=100 ms |
| 20 | 5 | 10 | 0.25 | 0.1 | 0.0 | neural lag=100 ms |
| 21 | 5 | 10 | 3.0 | 1.0 | 0.5 | 2-group test |
| 22 | 5 | 10 | 3.0 | 0.1 | 0.5 | nonstationary connection strengths |
| 23 | 5 | 10 | 3.0 | 0.1 | 0.5 | stationary connection strengths |
| 24 | 5 | 10 | 3.0 | 0.1 | 0.5 | only one strong external input |
| 25 | 5 | 5 | 3.0 | 1.0 | 0.5 | |
| 26 | 5 | 2.5 | 3.0 | 1.0 | 0.5 | |
| 27 | 5 | 2.5 | 3.0 | 0.1 | 0.5 | |
| 28 | 5 | 5 | 3.0 | 0.1 | 0.5 | |

**Table 1**: All the parameters included in the benchmark synthetic datasets (Smith et al., 2011)

**Supplementary material 3** Complex cumulants, a 2-node simulation in polar coordinates

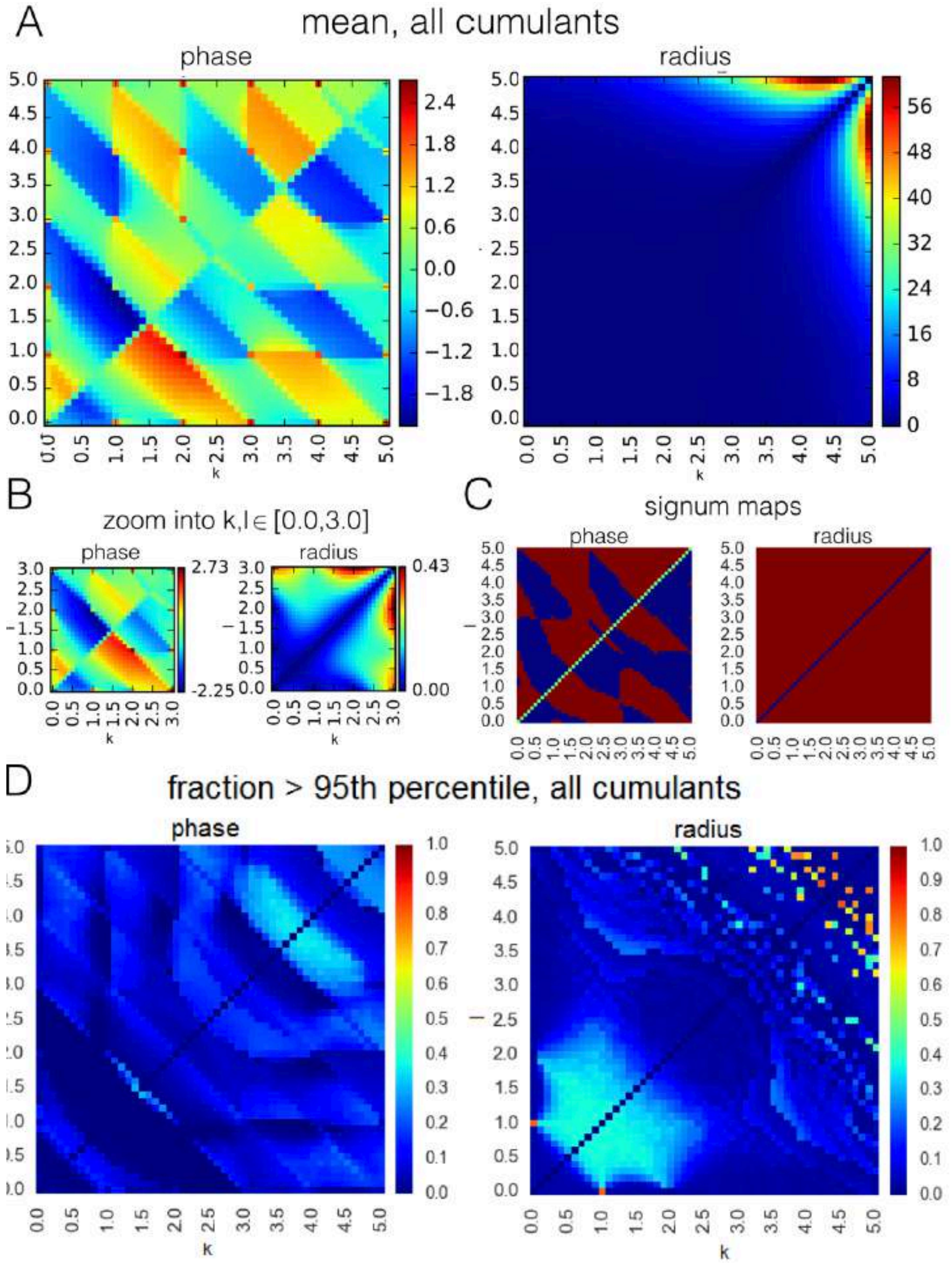

**Figure 9** Cumulants in polar coordinates. A: mean values for all cumulants, over 1000 simulations. The radius and the phase are computed with respect to point (0, 0), otherwise the (anti-)symmetry would be lost. Since cumulants are antisymmetric with respect to indexing k, l, the heatmaps for the phase are antisymmetric. The radius is always positive. B: zoom into a smaller range of [0.0; 3.0]. C: signum of the cumulants. Red: positive. Blue: negative. Green: zero. D: Discriminative power for all cumulants in range k, l in [0:0; 5:0], in polar coordinates (in the ideal case of a very long BOLD time series and no background neuronal noise). Cumulants in polar coordinates are less informative. Phase is uninformative since the discriminative power is relatively high only in the high-moment regime - and high moments are

**Supplementary material 4** Success rate for all the cumulants in the synthetic benchmark datasets

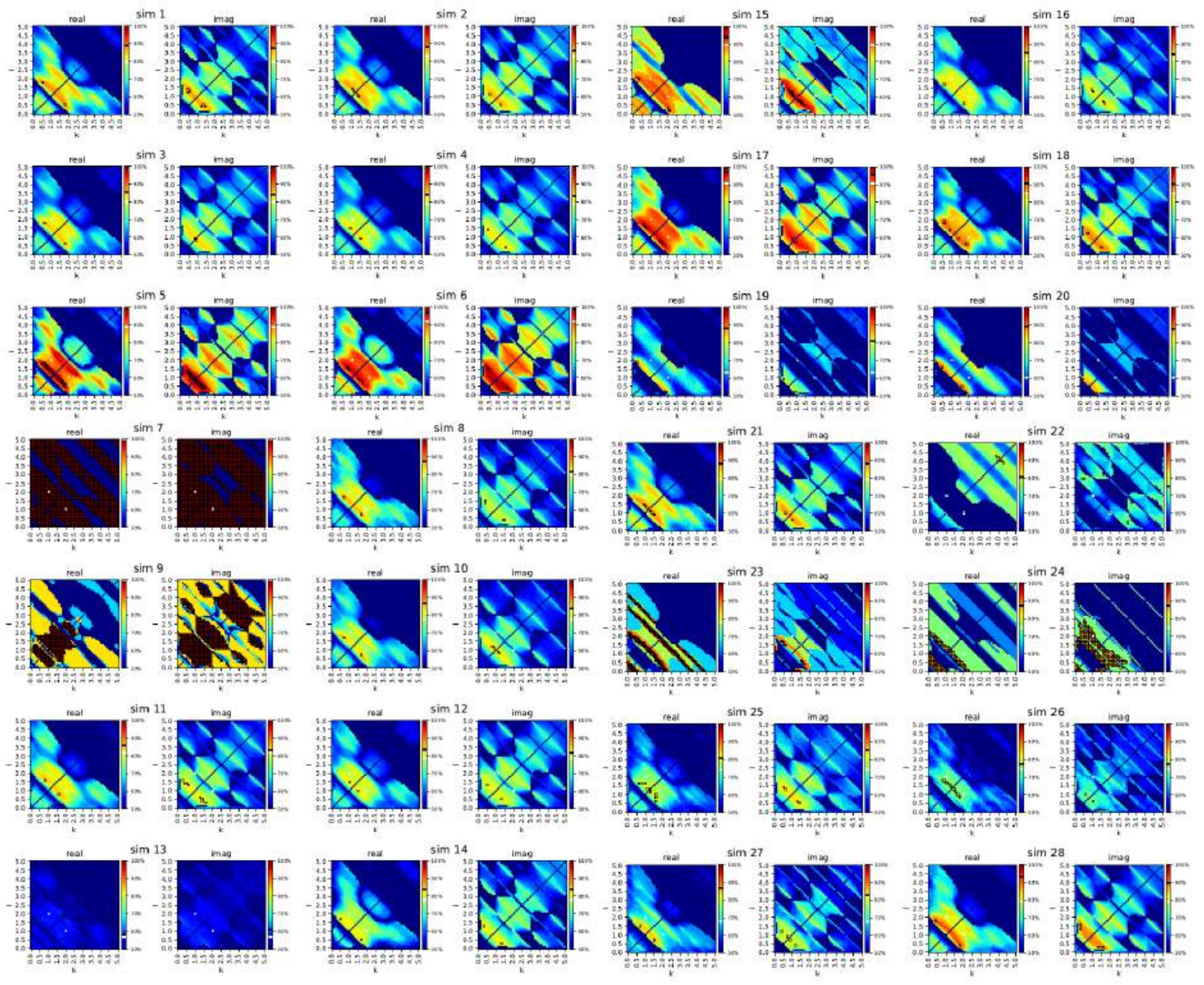

**Figure 10** Success rate for all the individual cumulants, for the remaining 27 simulations at benchmark datasets. In every case, there are cumulants performing better than the cumulant (2,1). Simulation no 7 is particularly convenient for effective connectivity methods whereas simulation no 13 is particularly challenging.

In Fig. 10, results for the remaining 27 simulations from Smith el al., 2011 are presented. For all the methods, simulation no 7 is particularly convenient for effective connectivity methods as it contains the most convenient choice of parameters: the network is small (N=5) and the simulation is unrealistically long (T = 250 [min]). As discussed in Smith et al., 2011, simulation length is one of the crucial factors influencing methods for effective connectivity. Simulation 13 is particularly hard on the other hand. In this simulation, each adjacency matrix additionally contains negative (inhibitory) backwards connections for a randomly chosen half of all connections (connectivity strength for new, backwards connections is sampled from the same distribution as the original connections). With inhibitory connections confounding the influence of exhibitory connections, none of the tested methods achieved mean performance higher than 60% success rate.

**Supplementary material 5** Comparison between methods on all the synthetic benchmark datasets

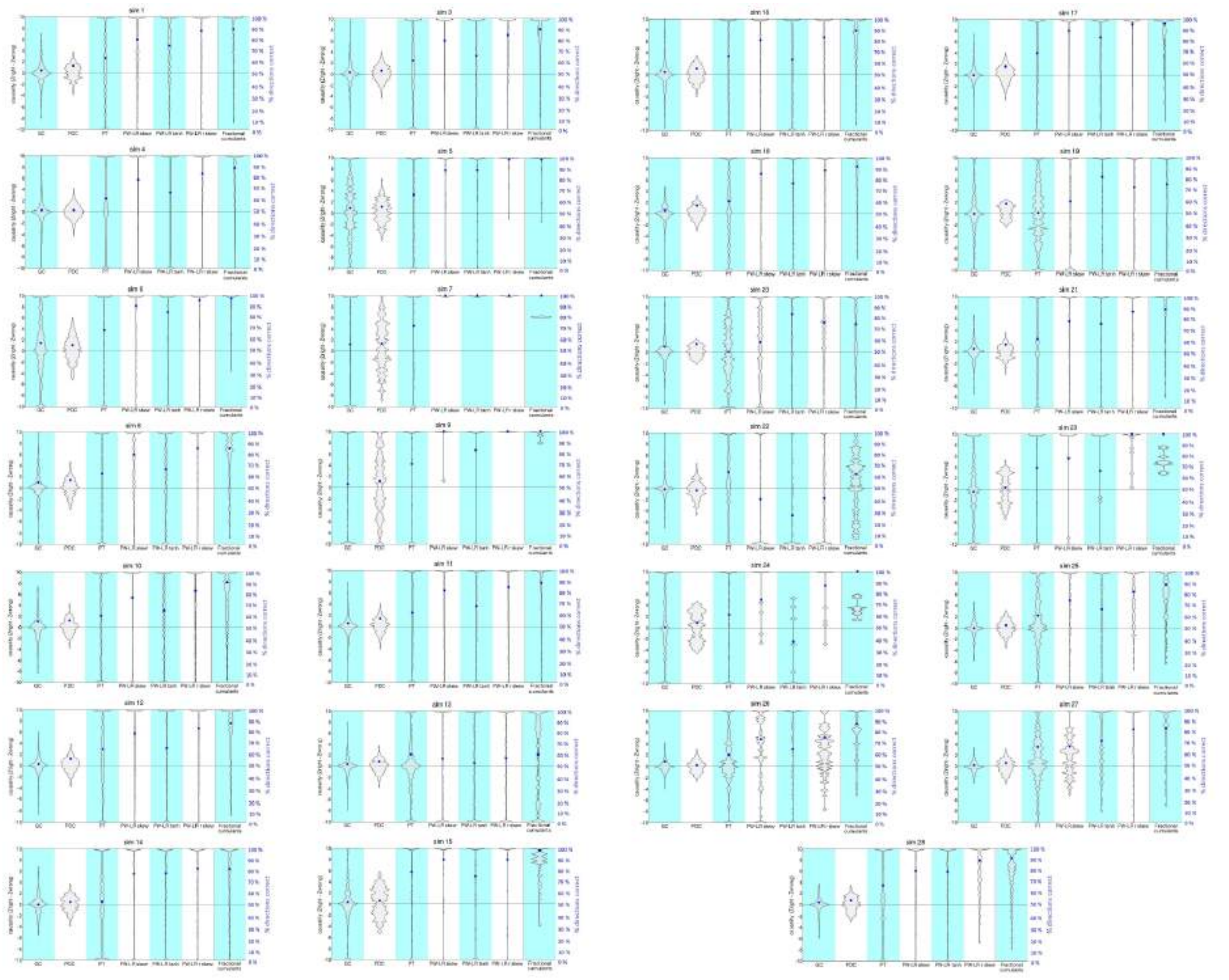

**Figure 11** Overall success rate of Fractional cumulants against 'PW-LR' methods. In most of the simulations, we achieved a slight improvement with respect to the main competitor, 'PW-LR r skew'. In simulations 7, 8, 9, 14, 17, 20 and 23, the performance are roughly the same.

**Supplementary material 6** ‘Net connectivity’: Cumulant maps for a simple two-node system, in case of a uni- versus bidirectional connection

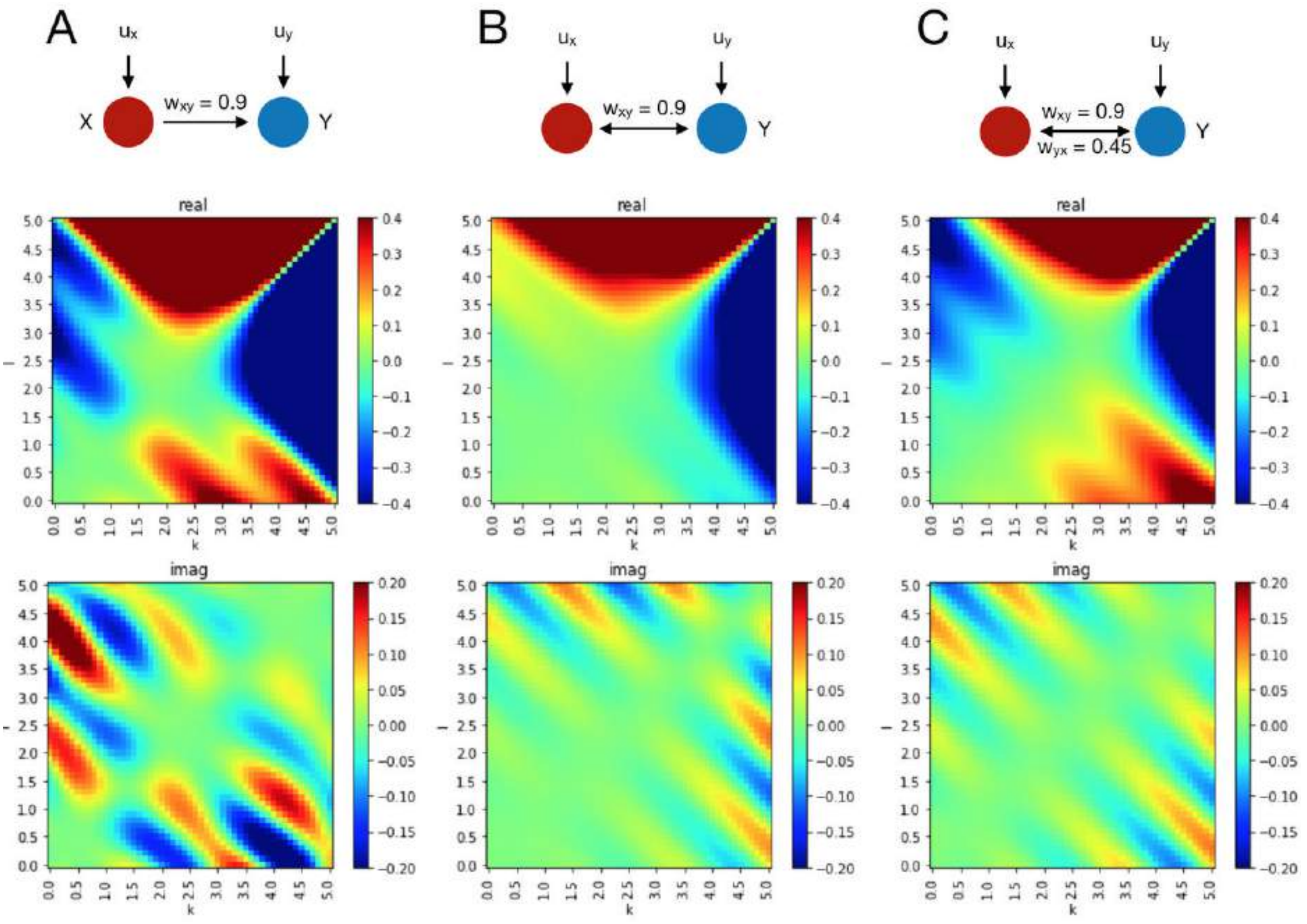

**Figure 12** Net connectivity effect. Simulation of a 10 [min] dynamics of a two-node linear system, Eq. 5, Fs = 200 [Hz], σ = 0, binary inputs simulated as stochastic trains of on- and off-states as described in the Methods section. A: unidirectional connection with connection weight w = 0.9, B: bidirectional connection with weights $w_{XY} = w_{YX} = 0.9$, C: asymmetric bidirectional connection with weights $w_{XY} = 0.9$, $w_{YX} = 0.45$. When the connection is perfectly symmetrical, B, the amplitudes of cumulants will be *weaker* than in case of an asymmetric, unidirectional connection, A. In case of bidirectional connection with *asymmetric weights* though, C, the cumulant maps will give more pronounced effect than in case of a perfectly bidirectional connection, B and weaker than in case of perfectly unidirectional connection, A. As a conclusion, what the cumulant maps reflect is the *difference* between two connectivity strengths X → Y and Y → X.

In Fig. 12, results of a simple simulation of a two-node linear system (simulated with ordinary differential equations, Eq. 5, 10 [min] simulation with Fs = 200 [Hz], σ = 0, binary inputs sim-ulated as stochastic trains of on- and off- states as described in the Methods section) are presented. We simulated the same system with the same set of inputs in three scenarios: (A) unidirectional connection with connection weight w = 0.9, (B) bidirectional connection with weights $w_{XY} = w_{YX} = 0.9$, (C) asymmetric bidirectional connection with weights $w_{XY} = 0.9$, $w_{YX} = 0.45$.

The results demonstrate that, in case the connection is perfectly symmetrical, B, the amplitudes of cumulants will be *weaker* than in case of an asymmetric, unidirectional connection, A. In case of bidirectional connection with *asymmetric weights* though, C, the cumulant maps will give more pronounced effect than in case of a perfectly bidirectional connection, B (as the system of less symmetrical in case), but at the same time, weaker than in case of perfectly unidirectional connection, A. As a conclusion, what the cumulant maps reflect is the *difference* between two connectivity strengths X → Y and Y → X. We refer to this effect as *net connectivity*.